\documentclass[twocolumn]{aastex63}
\usepackage{natbib}
\usepackage{amsmath}
\usepackage{bm}

\begin{document}
	
\title{Precise Transit Photometry Using TESS: Updated Physical Properties for 28 Exoplanets Around Bright Stars}
	
\author[0000-0001-8018-0264]{Suman Saha}
\affiliation{Indian Institute of Astrophysics, II Block, Koramangala, Bengaluru, India}
\affiliation{Pondicherry University, R.V. Nagar, Kalapet, Puducherry, India}
	
\correspondingauthor{Suman Saha}
\email{suman.saha@iiap.res.in}

\accepted{for publication in The Astrophysical Journal Supplement Series}

\begin{abstract}

The TESS follow-up of a large number of known transiting exoplanets provide unique opportunity to study their physical properties more precisely. Being a space-based telescope, the TESS observations are devoid of any noise component resulting from the interference of Earth's atmosphere. TESS also provides a better probability to observe subsequent transit events owing to its longer uninterrupted time-series observations compared to the ground-based telescopes. Especially, for the exoplanets around bright host-stars, TESS time-series observations provides high SNR lightcurves, which can be used for higher precision studies for these exoplanets. In this work, I have studied the TESS transit photometric follow-up observations of 28 exoplanets around bright stars with $V_{mag}\le$10. The already high SNR lightcurves from TESS have been further processed with a critical noise treatment algorithm, using the wavelet denoising and the Gaussian-process regression techniques, to effectively reduce the noise components both correlated and uncorrelated in time, which were then used to estimate the physical properties of these exoplanets. The study has resulted in very precise values for the physical properties of the target exoplanets, with the improvements in precision being significant for most of the cases compared to the previous studies. Also, since a comparatively large number of transit lightcurves from TESS observations were used to estimate these physical properties for each of the target exoplanets, which removes any bias due to the lack of sufficient datasets, these updated physical properties can be considered extremely accurate and reliable for future studies.

\end{abstract}


\section{Introduction}\label{sec1}

The Transiting Exoplanet Survey Satellite (TESS) \citep{2015JATIS...1a4003R} is a survey telescope to discover new exoplanets around the nearby bright stars. Over the span of this entire survey, TESS will cover a large portion ($>90\%$) of the sky. This also means that TESS automatically does follow-up observations of many of the previously known exoplanets. Most of the previously known exoplanets discovered by the ground-based survey missions have only been previously studies using the observations from small (sub-2m class) ground-based telescope, which are both affected by the noise components from the interference of Earth's atmosphere and limited by the less SNR in the observed lightcurves. This have resulted in large uncertainties in the currently known physical properties of these exoplanets. However, TESS being a space-based instrument, provides observations not affected by the Earth's atmosphere. Especially for the nearby bright stars, the TESS lightcurves have reasonably high SNR. This gives the unique opportunity to conduct the follow-up studies of the transiting exoplanets around the nearby bright stars, which can give more precise and accurate estimation of their physical properties.

\begin{deluxetable*}{lCCC}
	\tablecaption{Targets and observational details \label{tab:tab1}}
	\tablewidth{0pt}
	\tablehead{\colhead{Target Name} & \colhead{Host star $V_{mag}$} & \colhead{Sector} & \colhead{No. of full transits}}
	\startdata
	KELT-2 A b & 8.68 & 43-45 & 16\\
	KELT-3 b & 9.82 & 21,48 & 17\\
	KELT-4 A b & 9.98 & 48 & 7\\
	KELT-11 b & 8.04 & 9 & 5\\
	KELT-17 b & 9.23 & 44-46 & 22\\
	KELT-19 A b & 9.86 & 7,33 & 4\\
	KELT-20 b & 7.59 & 14,40,41,54 & 27\\
	KELT-24 b & 8.34 & 14,20,21,40,41,47,48 & 32\\
	HAT-P-1 b & 9.83 & 56 & 5\\
	HAT-P-2 b & 8.72 & 24,25,51,52 & 16\\
	HAT-P-11 b & 9.46 & 14,15,41,54-56 & 31\\
	HAT-P-22 b & 9.76 & 21,48 & 14\\
	HAT-P-69 b & 9.77 & 7,34 & 4\\
	HAT-P-70 b & 9.47 & 5,32 & 9\\
	MASCARA-4 b & 8.19 & 10,11,36,38 & 17\\
	XO-3 b & 9.85 & 19 & 6\\
	WASP-7 b & 9.5 & 27 & 4\\
	WASP-8 b & 9.79 & 2,29 & 4\\
	WASP-14 b & 9.75 & 50 & 6\\
	WASP-18 b & 9.28 & 2,3,29,30 & 91\\
	WASP-33 b & 8.14 & 18 & 16\\
	WASP-69 b & 9.87 & 55,81 & 4\\
	WASP-76 b & 9.52 & 30,42,43 & 33\\
	WASP-99 b & 9.48 & 3,29,30 & 12\\
	WASP-136 b & 9.97 & 29,42 & 7\\
	WASP-166 b & 9.35 & 8,35 & 6\\
	WASP-178 b & 9.95 & 11,38 & 7\\
	WASP-189 b & 6.6 & 51 & 3\\
	\enddata
\end{deluxetable*}

In this work, I have studied the transit photometric follow-up observations of 28 exoplanets around bright stars with $V_{mag}\le$10 from TESS, to estimate their physical properties with greater accuracy and precision compared to the previous studies. Being very bright stars, the TESS photometric lightcurves obtained for these targets are expected to have high SNR. Also, I have found that for most these targets, the currently known parameter values estimated in the previous studies have large uncertainties, as they were previously estimated from the ground-based transit observations. This motivates for a transit follow-up study using TESS observations, which could provide a better estimation of these physical properties.

One of the major factors that limit the capability of ground-based as well as space-based telescopes is the noise components in the signal, originating from various sources. Broadly, these noise components can be categorized into two types, noise components which are uncorrelated in time, and noise components which are correlated in time. The noise uncorrelated in time originates from several instrumental factors, outliers due to various astronomical phenomenons, and in the case of ground-based observations, the variability of Earth's atmosphere. On the other hand, the noise correlated in time originates due to stellar activity and pulsations, small-scale variability of the planet hosting stars, and instrumental effects. Previously, \cite{2019AJ....158...39C}, \cite{2021AJ....162...18S} and \cite{2021AJ....162..221S} have developed a critical noise treatment algorithm using the wavelet denoising and the Gaussian-process (GP) regression techniques to reduce the noise components both uncorrelated and correlated in time from the transit lightcurves. \cite{2021AJ....162..221S} have applied this algorithm to the transit photometric data from TESS and demonstrated its effectiveness in estimating the physical properties of the transiting exoplanets more precisely. I have used the same algorithm as was used in \cite{2021AJ....162..221S} to effectively deal with the noise components present in the TESS transit lightcurves analyzed in this work.

In section \ref{sec2}, I have discussed about the target selection and observations; in section \ref{sec3}, I have detailed about the data analysis and modeling techniques; and finally in section \ref{sec4}, I have discussed about the results obtained from this work.

\section{Target selection and observational data}\label{sec2}

For this study, I have selected those transiting exoplanets, which are orbiting around stars with $V_{mag}\le$10, and have TESS follow-up observational data from one or more than one sectors. The 28 exoplanets selected through this criteria are: KELT-2 b, KELT-3 b, KELT-4 A b, KELT-11 b, KELT-17 b, KELT-19 A b, KELT-20 b, KELT-24 b, HAT-P-1 b, HAT-P-2 b, HAT-P-11 b, HAT-P-22 b, HAT-P-69 b, HAT-P-70 b, MASCARA-4 b, XO-3 b, WASP-7 b, WASP-8 b, WASP-14 b, WASP-18 b, WASP-33 b, WASP-69 b, WASP-76 b, WASP-99 b, WASP-136 b, WASP-166 b, WASP-178 b and WASP-189 b. While most of these exoplanets are hot to warm Jupiters, HAT-P-11 b is a warm Neptune and WASP-166 is a warm Saturn.

The TESS PDCSAP observational light-curves \citep{2014PASP..126..100S, 101086667697, 2012PASP..124..985S, 2017ksci.rept....2J} of these targets was obtained from the public Mikulski Archive for Space Telescopes
(MAST)\footnote{https://mast.stsci.edu/portal/Mashup/Clients/Mast/Portal.html}. In Table \ref{tab:tab1}, I have listed the $V_{mag}$ of the host-stars (obtained from NASA Exoplanet Archive\footnote{https://exoplanetarchive.ipac.caltech.edu/}), the TESS sectors of observations and the number of full transits observed for each of the target exoplanets.

\section{Data Analysis and Modeling}\label{sec3}

The TESS lightcurves obtained from MAST for each of the targets from each sector spans over $\sim$ 27 days. I have identified the full transit observations in those lightcurves, and sliced them into smaller transit lightcurves, which were used for analysis. Only the full transit observations were considered in this study, as it removes the possibility of bias due to incomplete baseline. Also, the TESS observations for the exoplanets targeted in this study had sufficient number of full-transit observations to avoid any bias in the analysis due to insufficient datasets.

The transit lightcurves were then baseline corrected by modeling the out-of-transit sections with a first-order polynomial and subtracting it from the entire lightcurves. Baseline correction removes any large-scale correlated noise components in the signal either due to instrumental effects or the long-term stellar variability. I refrained from using a higher order polynomial for baseline correction, as it may induce unwanted distortions in the transit signal in the lightcurves, and also because the GP regression technique would be used at a later stage to remove any shorter-scale correlated noise components.

The lightcurves were then processed with the wavelet denoising technique \citep{Donoho1994IdealDI, 806084, WaveletDenoise2012, 2019AJ....158...39C, 2021AJ....162...18S, 2021AJ....162..221S} to reduce the time-uncorrelated fluctuations in the lightcurves. Unlike other smoothing techniques, like binning, the wavelet denoising technique uses wavelet transform to segregate the low amplitude noise components from the high amplitude signal, while preserving the valuable high frequency components arising from the transit event in the lightcurves. I have followed the same procedure for wavelet denoising as is given by \cite{2021AJ....162..221S}. The analysis uses PyWavelets \citep{Lee2019} python package for wavelet operations using the Symlet family of wavelets \citep{daubechies1988orthonormal}, which are the least asymmetric modified versions of the Daubechies wavelets \citep{daubechies1992ten, 1995ComPh...9..635R}. A single level of wavelet transform was used to avoid over-smoothing of the lightcurves, and the widely adopted universal thresholding law \citep{Donoho1994IdealDI} was used to estimate the threshold values for the noise level.

To reduce the correlated noise components in the transit lightcurves, I have used the GP regression technique \citep{2006gpml.book.....R, 2015ApJ...810L..23J, 2019MNRAS.489.5764P, 2020AA...634A..75B, 2019AJ....158...39C, 2021AJ....162...18S, 2021AJ....162..221S}. I have used the same procedure for GP regression as is discussed in \cite{2021AJ....162..221S}. GP regression is used to model the noise components correlated in time in the lightcurves while simultaneously modeling for the transit signal. While applying GP regression, I have used the Matérn class covariance function with the parameter of covariance, $\nu = 3/2$, and two free parameters, i.e. the signal standard deviation $\alpha$ and the characteristic length scale $\tau$, which are used as GP regression model parameters.

For modeling the transit signature in the lightcurves, the analytical transit formalism by \citet{2002ApJ...580L.171M} was used, which also incorporates the limb darkening effect using the quadratic limb darkening law. The Markov-chain Monte-Carlo (MCMC) sampling technique was used to simultaneously model the transit lightcurves for transit signature and the correlated noise, which incorporated the Metropolis-Hastings algorithm \citep{1970Bimka..57...97H}.

The directly estimated parameters from modeling the transit lightcurves, $b$, $R_\star/a$ and $R_p/R_\star$, were used along with the radial velocity and stellar parameters from the previous studies to derive other physical properties for the target exoplanets. The previous studies, from which the radial velocity and stellar parameters were adopted are: \cite{2019AJ....158..138S}, \cite{2012ApJ...756L..39B}, \cite{2017AJ....153..136S}, \cite{2013ApJ...773...64P}, \cite{2016AJ....151...45E}, \cite{2017AJ....153..215P}, \cite{2016AJ....152..136Z}. \cite{2018AJ....155...35S}, \cite{2018A&A...612A..57T}, \cite{2017AJ....154..194L}, \cite{2019A&A...631A..76H}, \cite{2019AJ....158..197R}, \cite{2014MNRAS.437...46N}, \cite{2018AJ....156..213M}, \cite{2014A&A...570A..80T}, \cite{2018AJ....155..255Y}, \cite{2018A&A...613A..41M}, \cite{2017A&A...602A.107B}, \cite{2019AJ....158..141Z}, \cite{2020A&A...635A..60D}, \cite{2012MNRAS.426.1291S}, \cite{2014ApJ...785..126K}, \cite{2020AA...636A..98C}, \cite{2015A&A...578L...4L}, \cite{2016A&A...585A.126W}, \cite{2014MNRAS.440.1982H}, \cite{2017A&A...599A...3L}, \cite{2019MNRAS.488.3067H}, \cite{2020AJ....160..111R}, \cite{2019MNRAS.490.1479H}, \cite{2020AA...643A..94L}. The mid-transit times estimated from modeling the transit lightcurves were used to estimate the transit ephemeris parameters, $T_0$ and $P$.

\begin{figure*}
	\centering
	\includegraphics[width=0.9\linewidth]{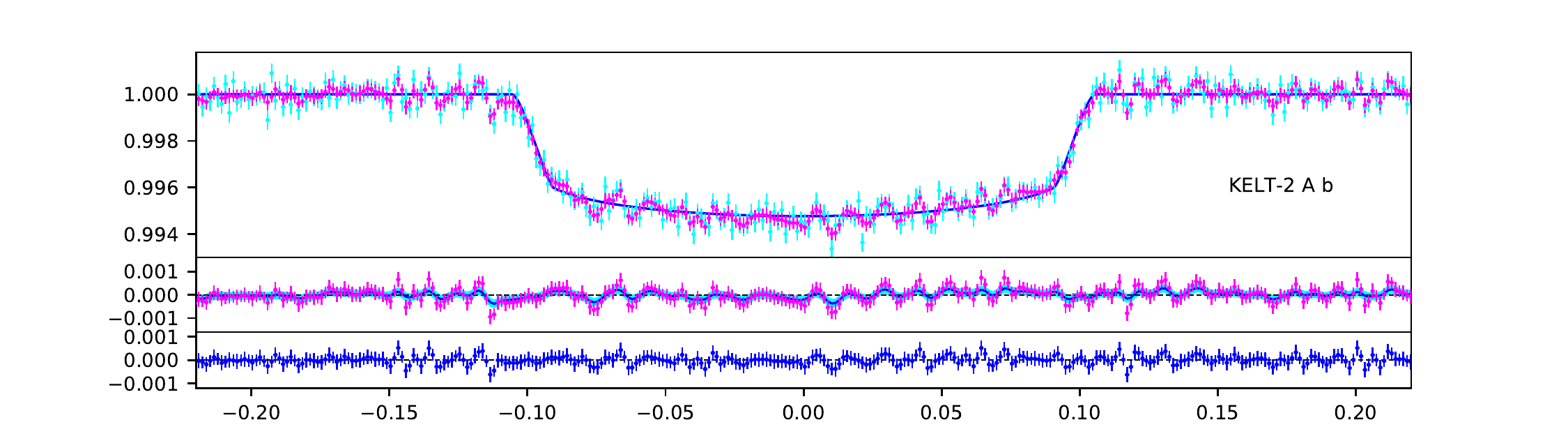}
	\includegraphics[width=0.9\linewidth]{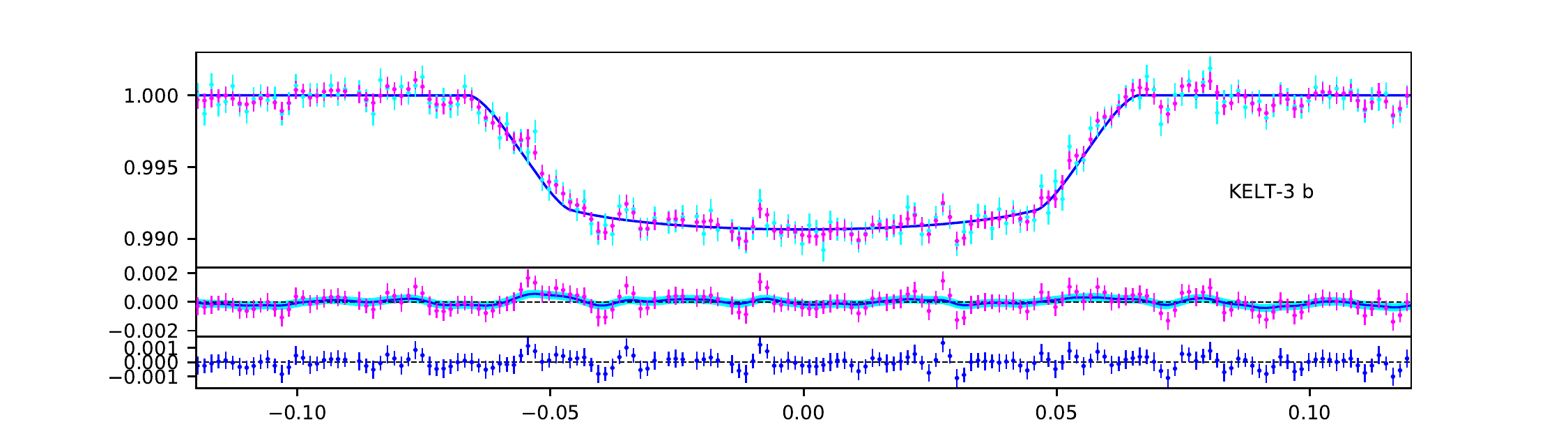}
	\includegraphics[width=0.9\linewidth]{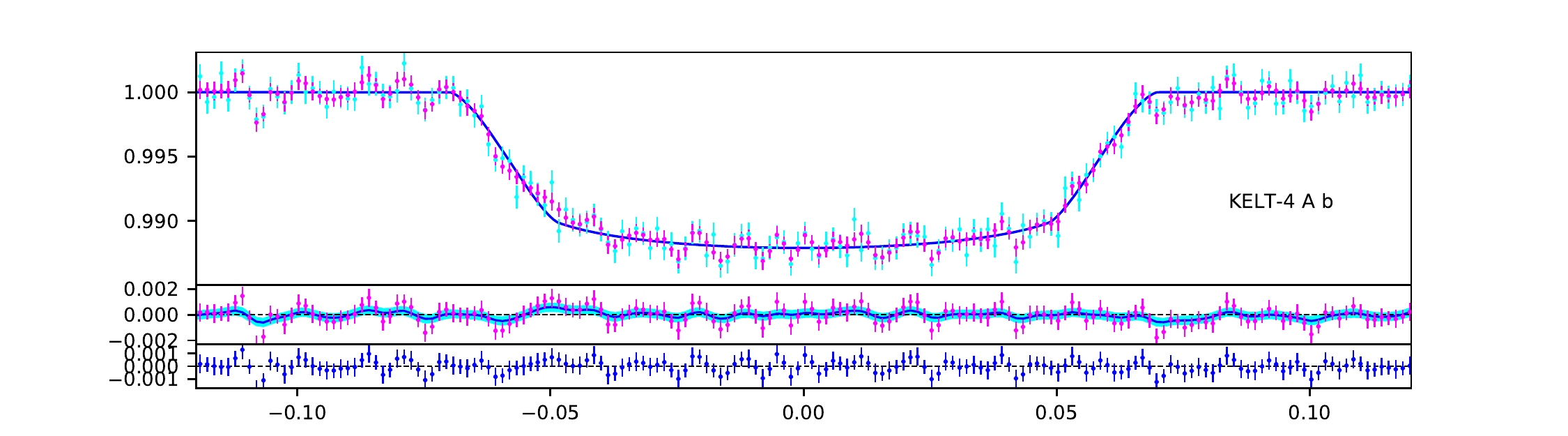}
	\includegraphics[width=0.9\linewidth]{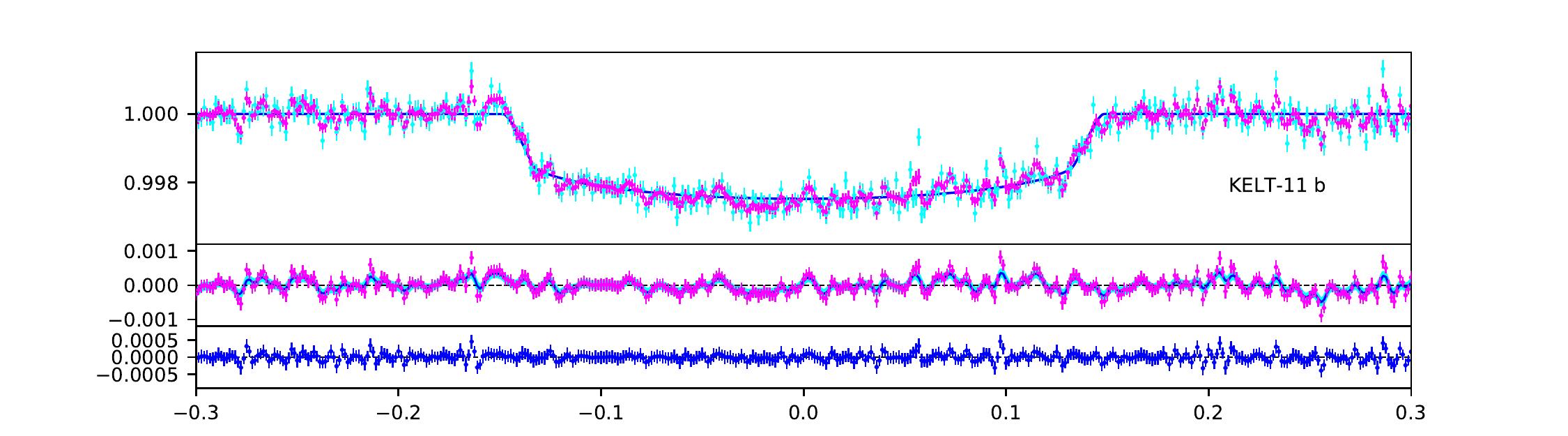}
	\includegraphics[width=0.9\linewidth]{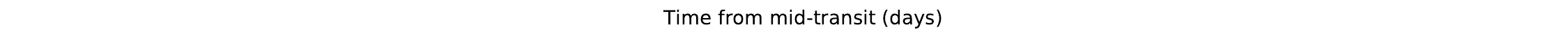}
	\caption{Observed and best-fit model light-curves (one transit event) for KELT-2 A b, KELT-3 b, KELT-4 A b and KELT-11 b. For each observed transit, Top: the unprocessed light-curve (cyan), light-curve after wavelet denoising (magenta), the best-fit transit model (blue). Middle: the residual after modeling without GP regression (magenta), the mean (blue) and 1-$\sigma$ interval (cyan) of the best-fit GP regression model. Bottom: mean residual flux (blue).}
	\label{fig:fig1}
\end{figure*}

\begin{figure*}
	\centering
	\includegraphics[width=0.9\linewidth]{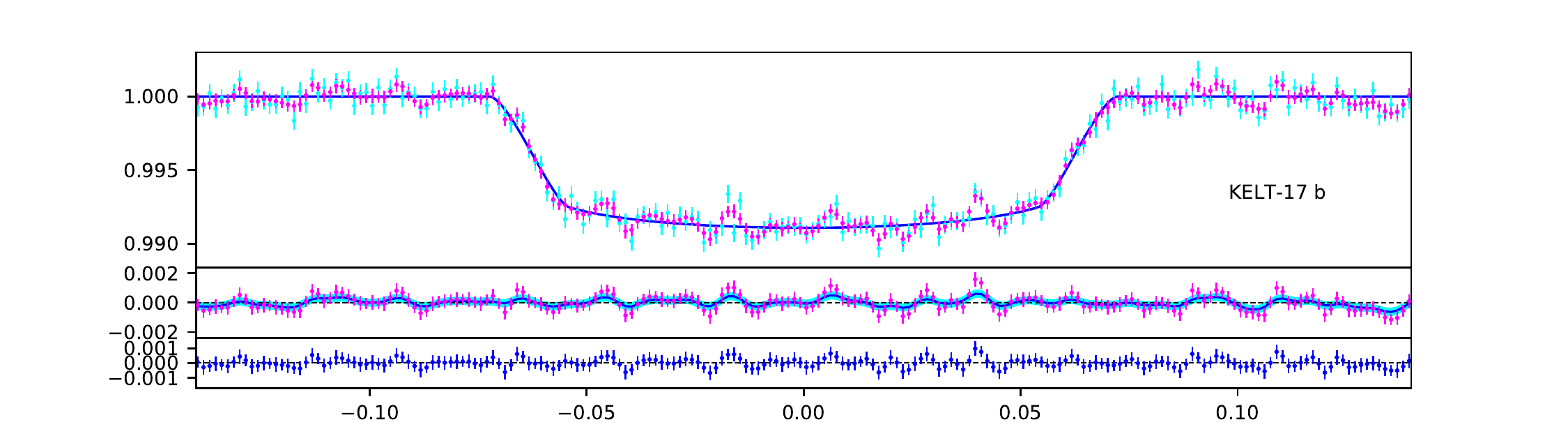}
	\includegraphics[width=0.9\linewidth]{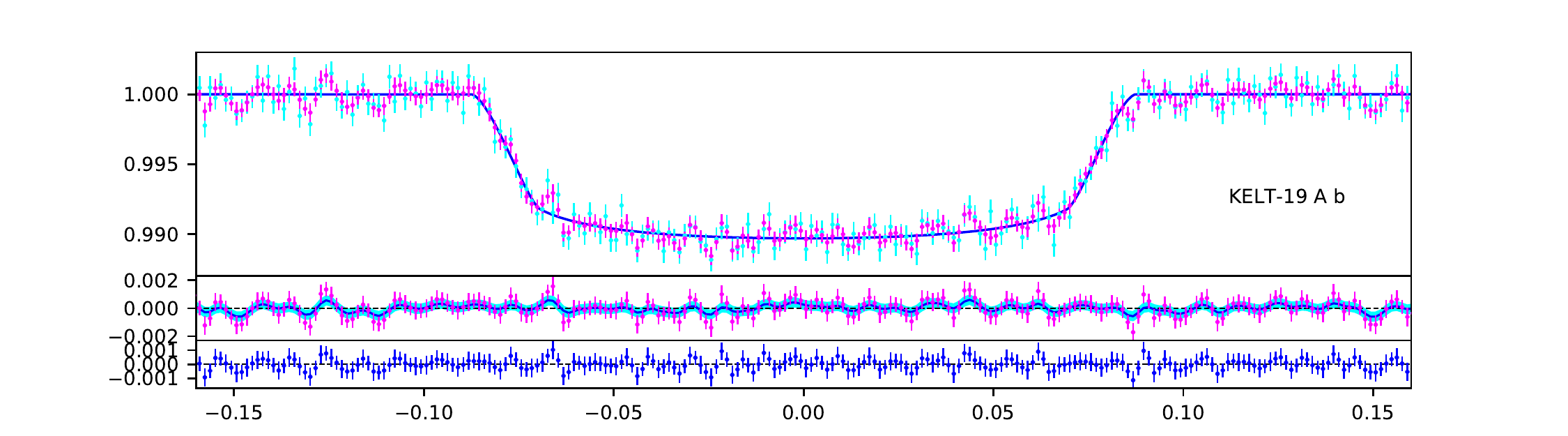}
	\includegraphics[width=0.9\linewidth]{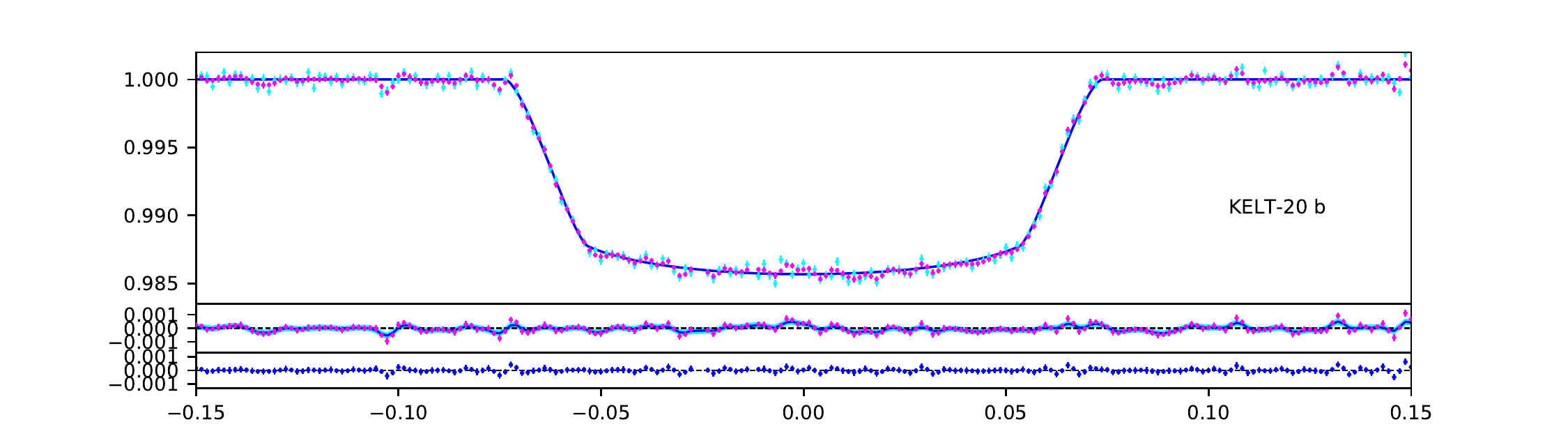}
	\includegraphics[width=0.9\linewidth]{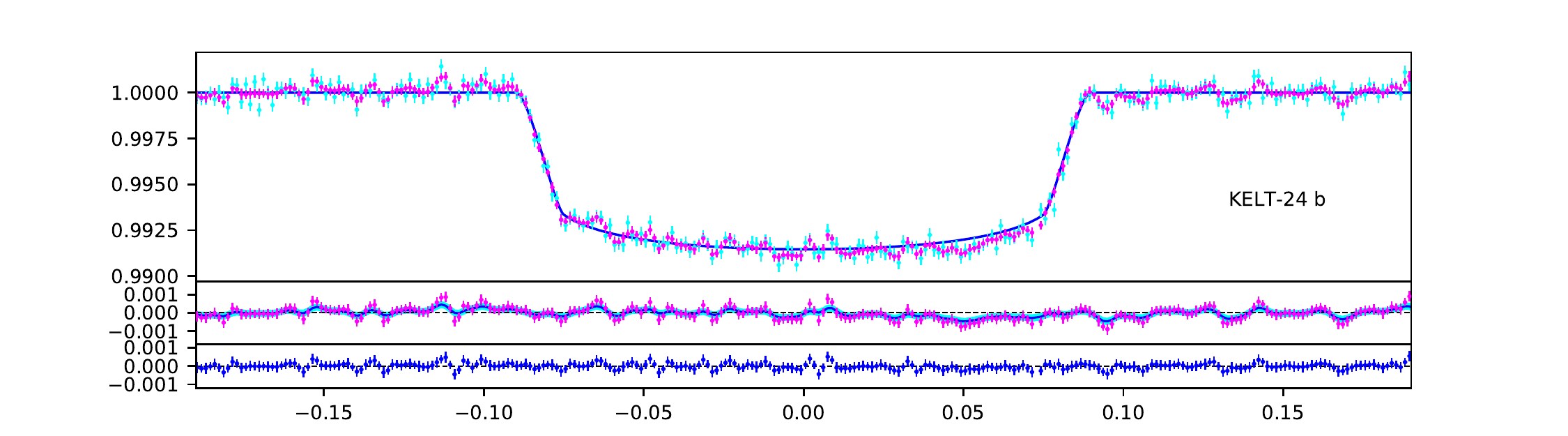}
	\includegraphics[width=0.9\linewidth]{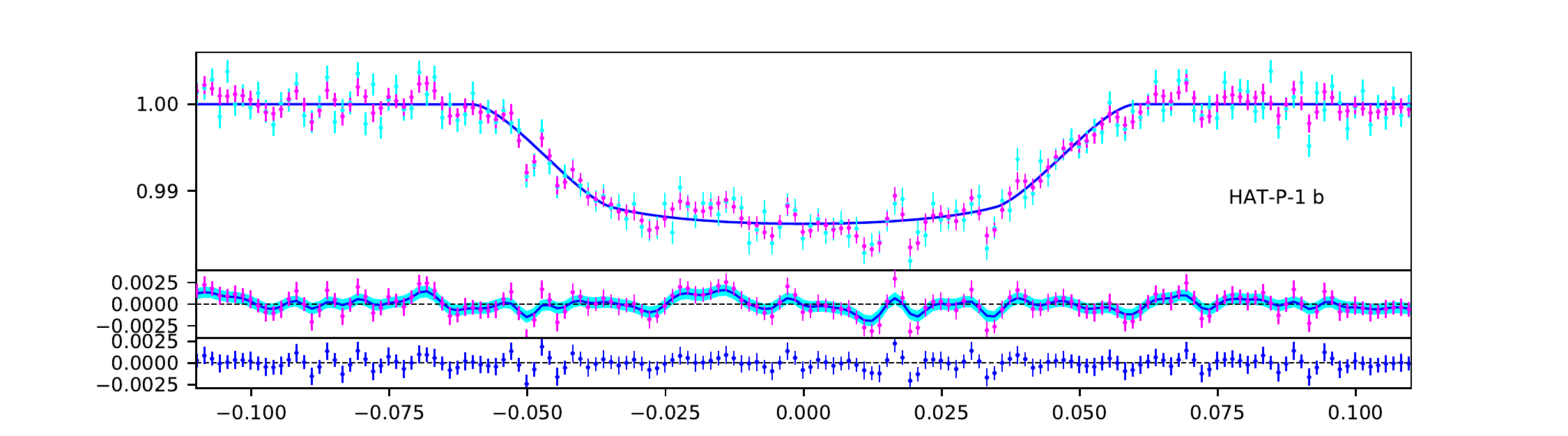}
	\includegraphics[width=0.9\linewidth]{f0.pdf}
	\caption{Same as Figure \ref{fig:fig1}, but for KELT-17 b, KELT-19 A b, KELT-20 b, KELT-24 b and HAT-P-1 b.}
	\label{fig:fig2}
\end{figure*}

\begin{figure*}
	\centering
	\includegraphics[width=0.9\linewidth]{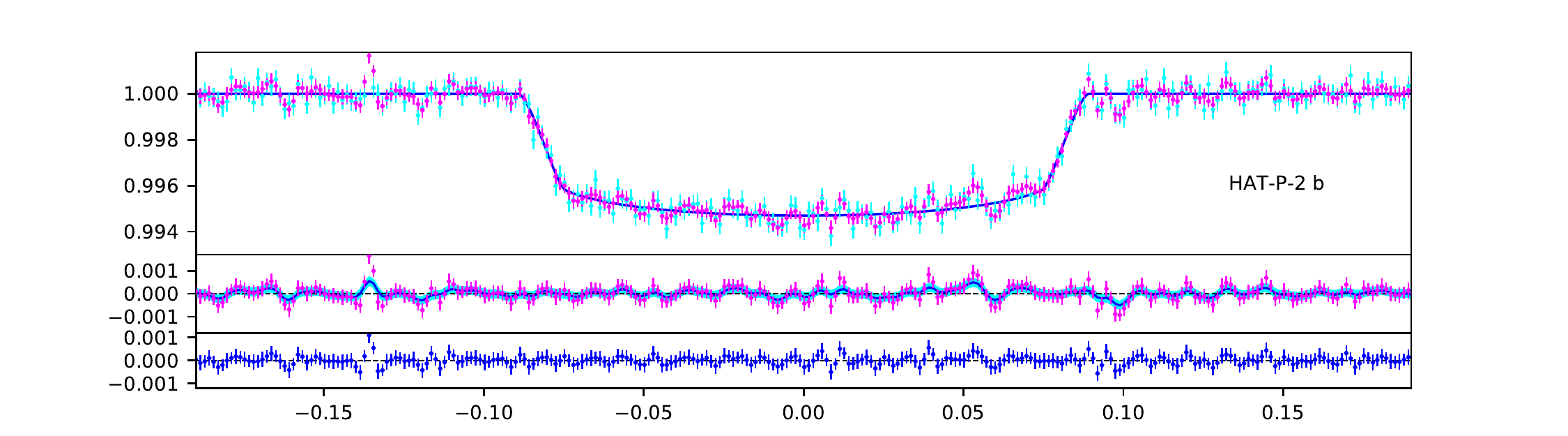}
	\includegraphics[width=0.9\linewidth]{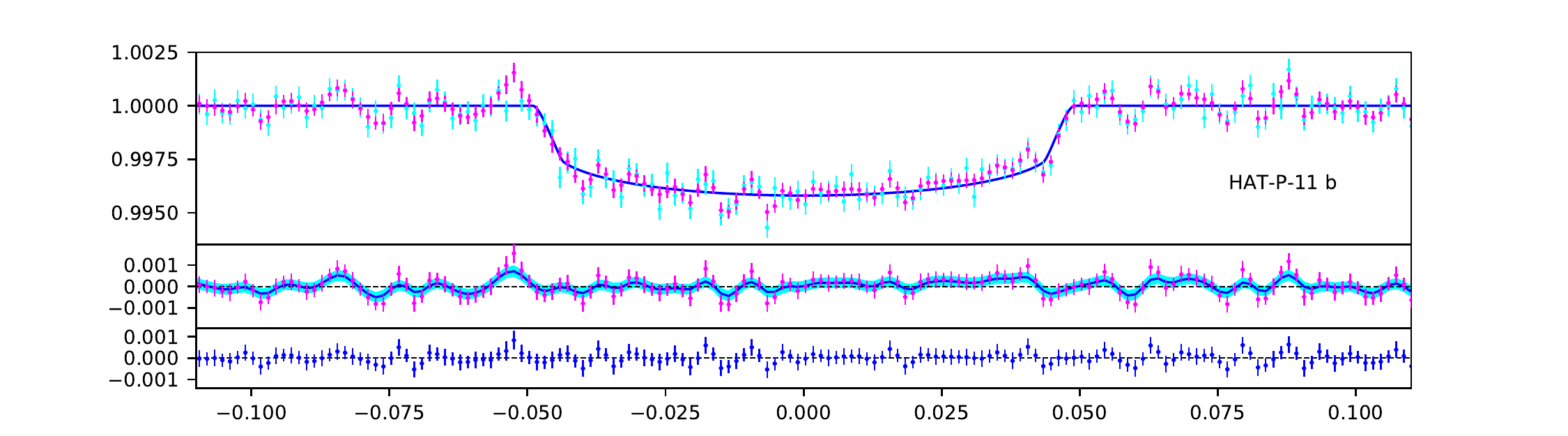}
	\includegraphics[width=0.9\linewidth]{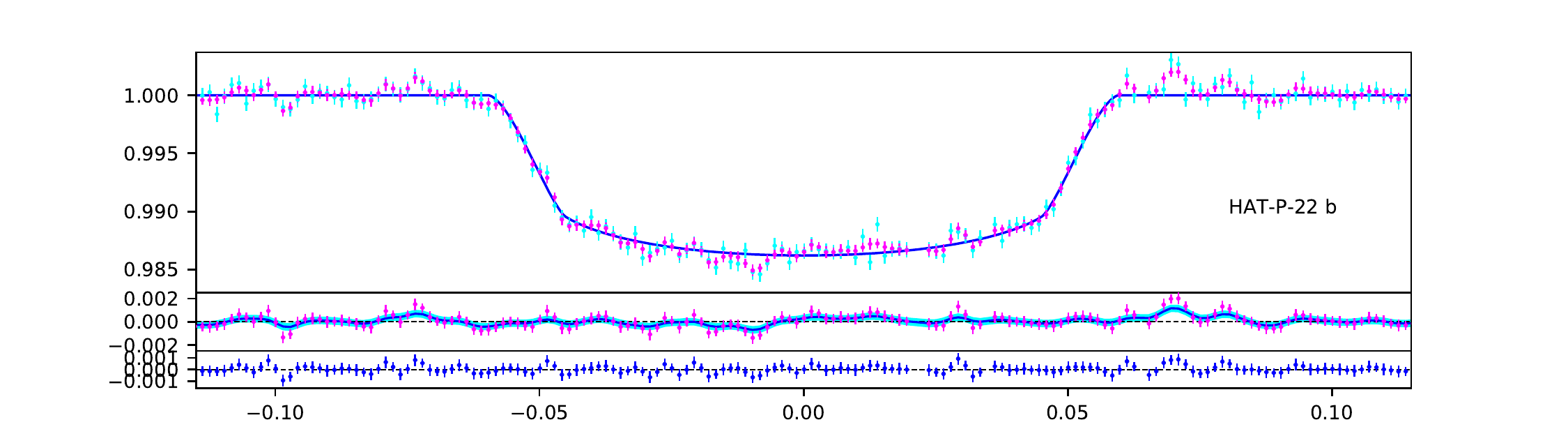}
	\includegraphics[width=0.9\linewidth]{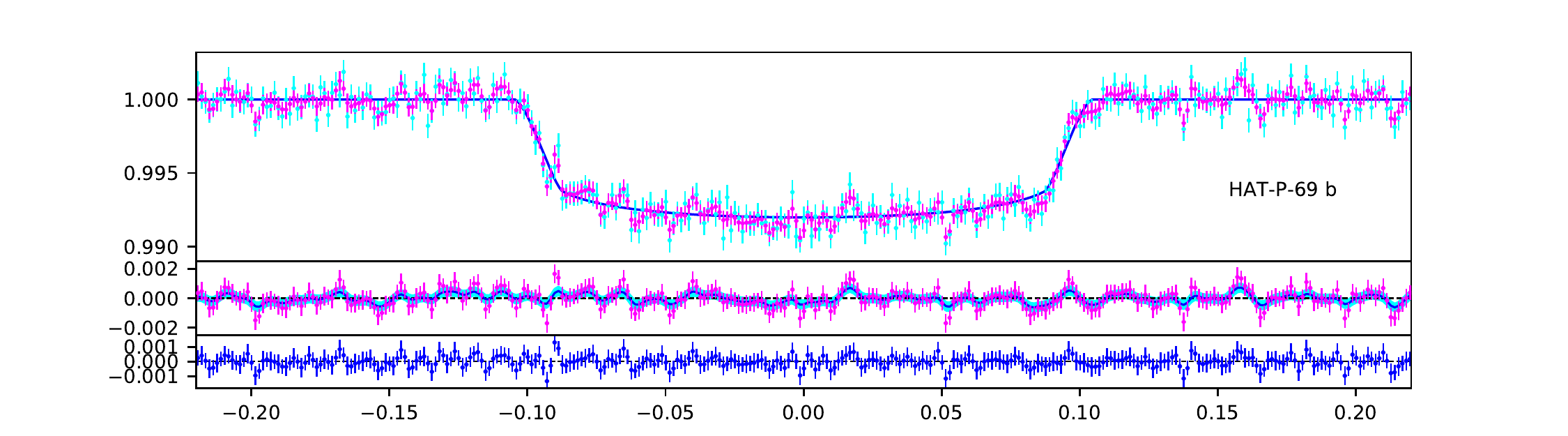}
	\includegraphics[width=0.9\linewidth]{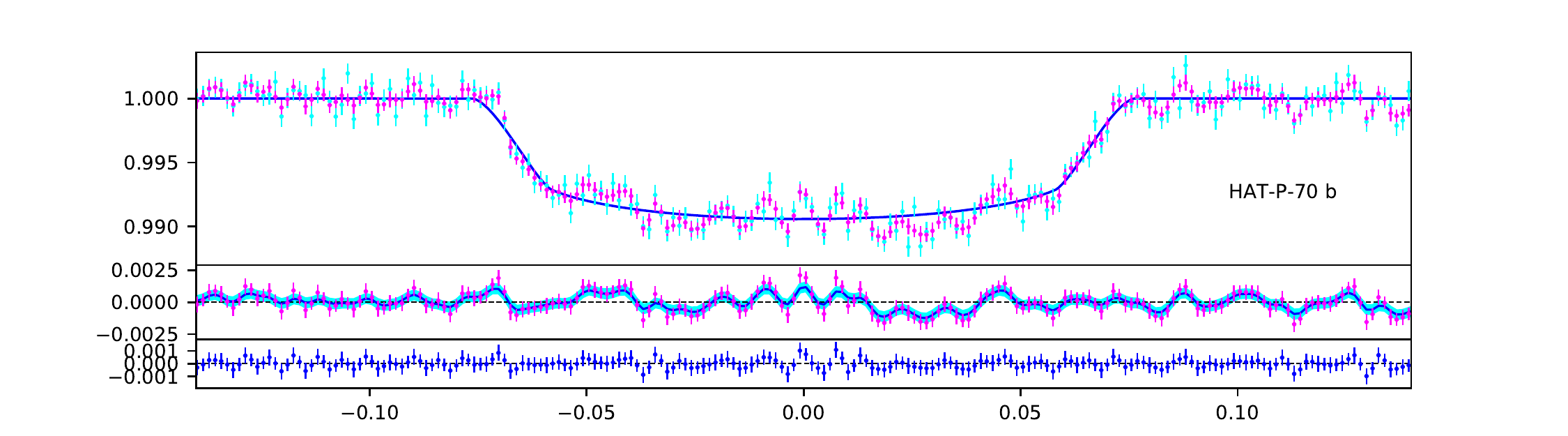}
	\includegraphics[width=0.9\linewidth]{f0.pdf}
	\caption{Same as Figure \ref{fig:fig1}, but for HAT-P-2 b, HAT-P-11 b, HAT-P-22 b, HAT-P-69 b and HAT-P-70 b.}
	\label{fig:fig3}
\end{figure*}

\begin{figure*}
	\centering
	\includegraphics[width=0.9\linewidth]{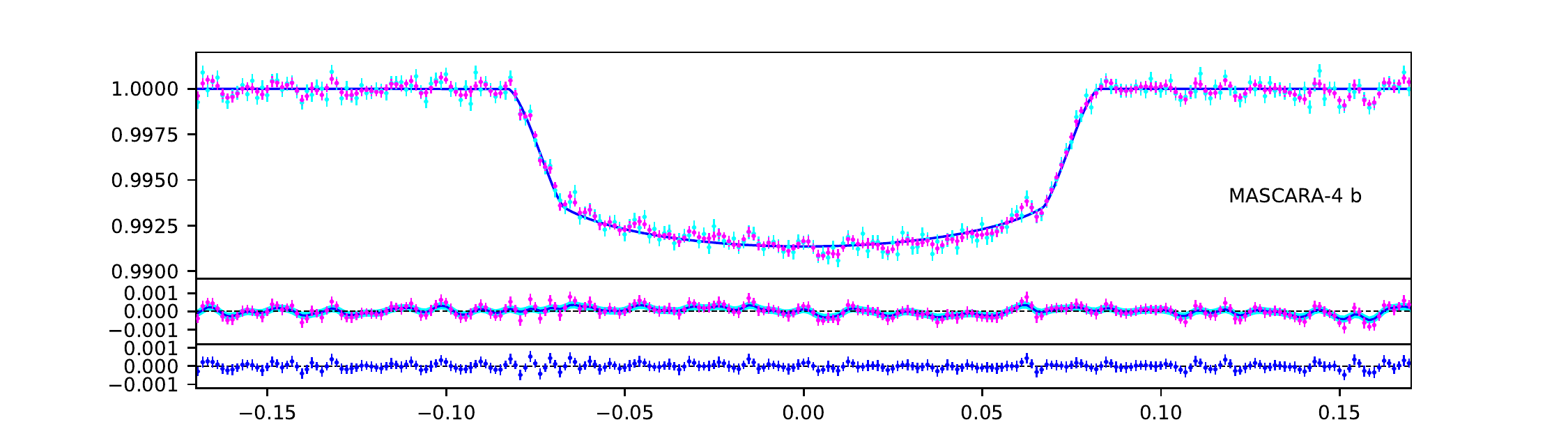}
	\includegraphics[width=0.9\linewidth]{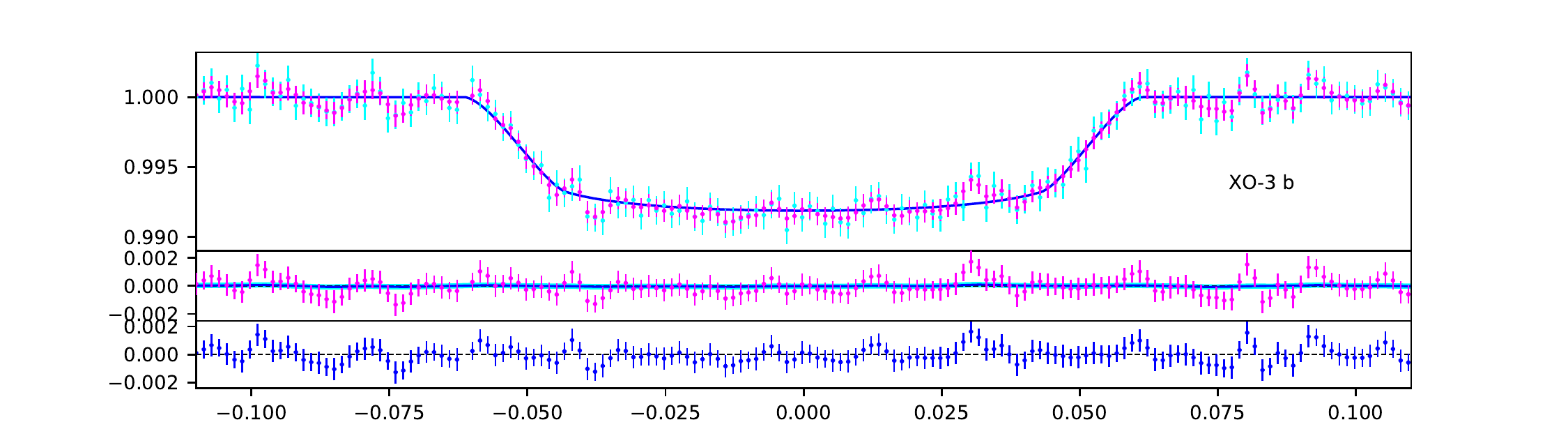}
	\includegraphics[width=0.9\linewidth]{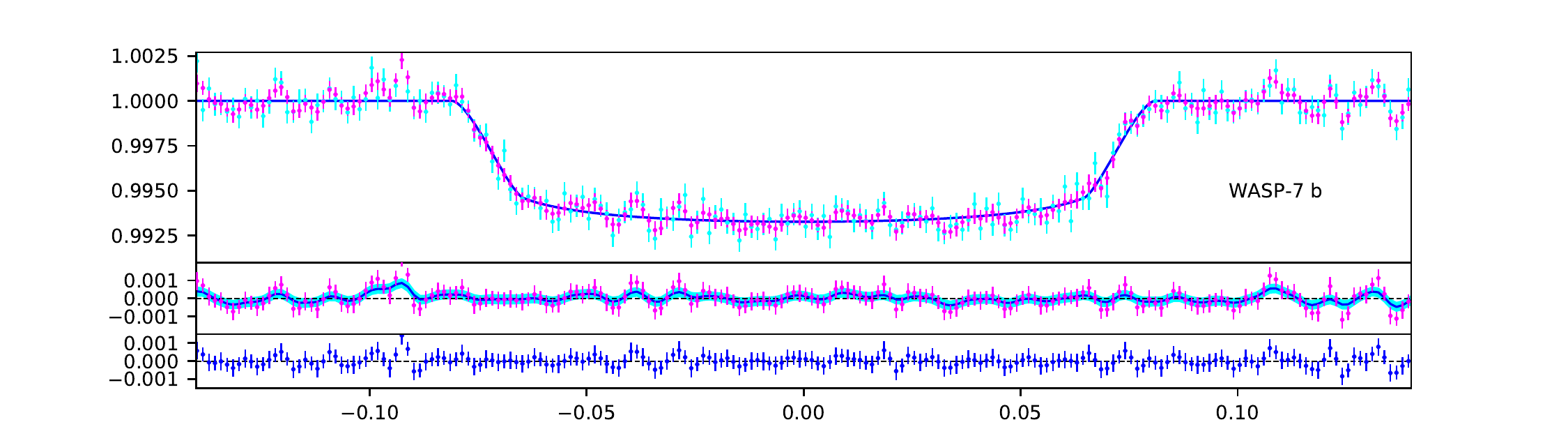}
	\includegraphics[width=0.9\linewidth]{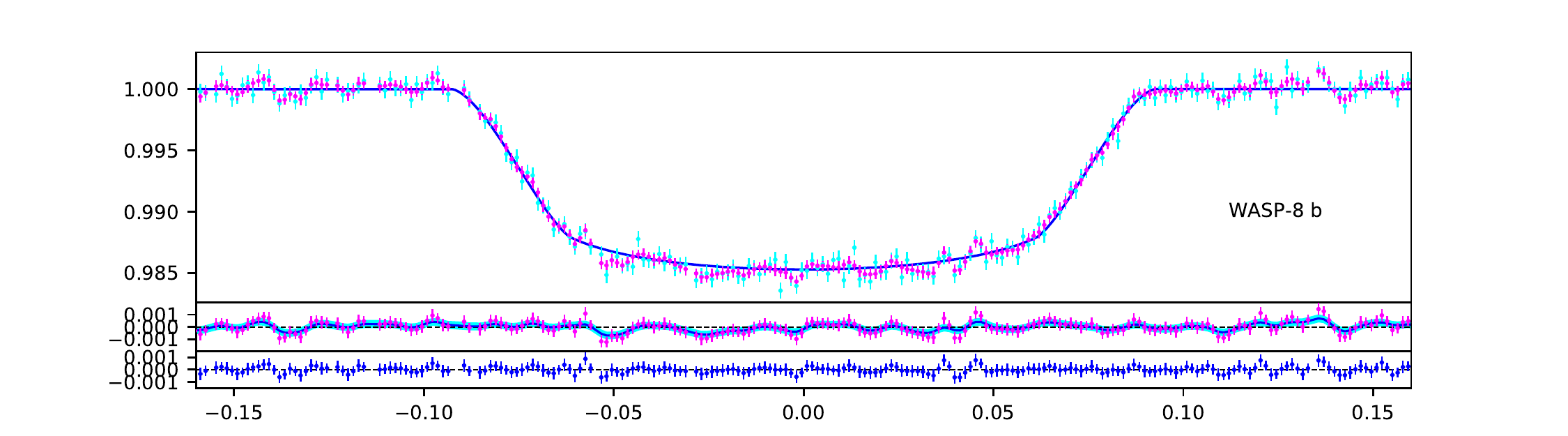}
	\includegraphics[width=0.9\linewidth]{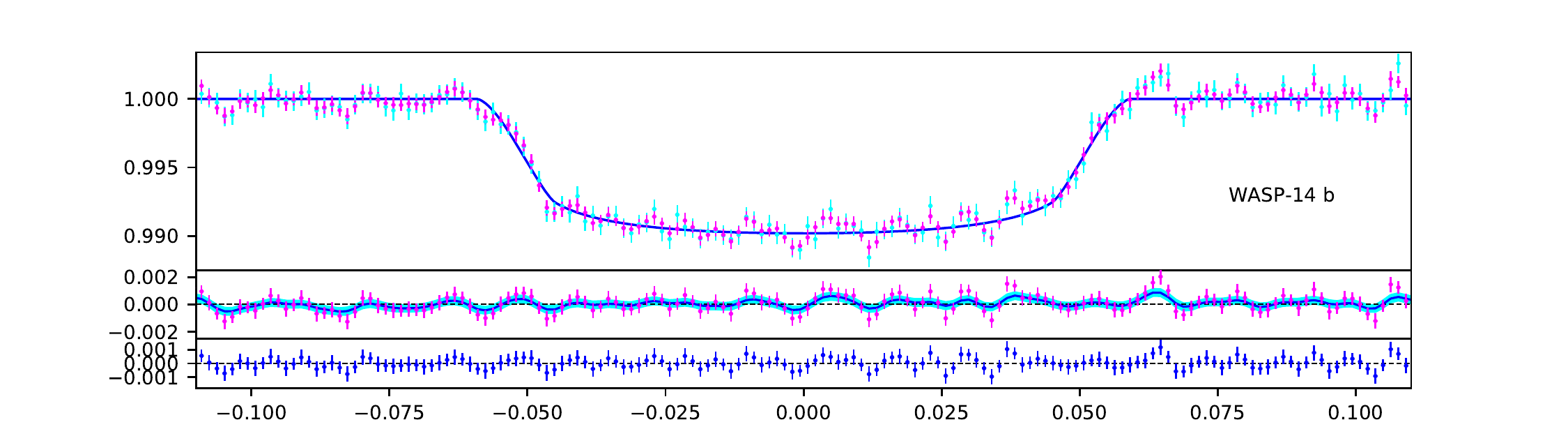}
	\includegraphics[width=0.9\linewidth]{f0.pdf}
	\caption{Same as Figure \ref{fig:fig1}, but for MASCARA-4 b, XO-3 b, WASP-7 b, WASP-8 b and WASP-14 b.}
	\label{fig:fig4}
\end{figure*}

\begin{figure*}
	\centering
	\includegraphics[width=0.9\linewidth]{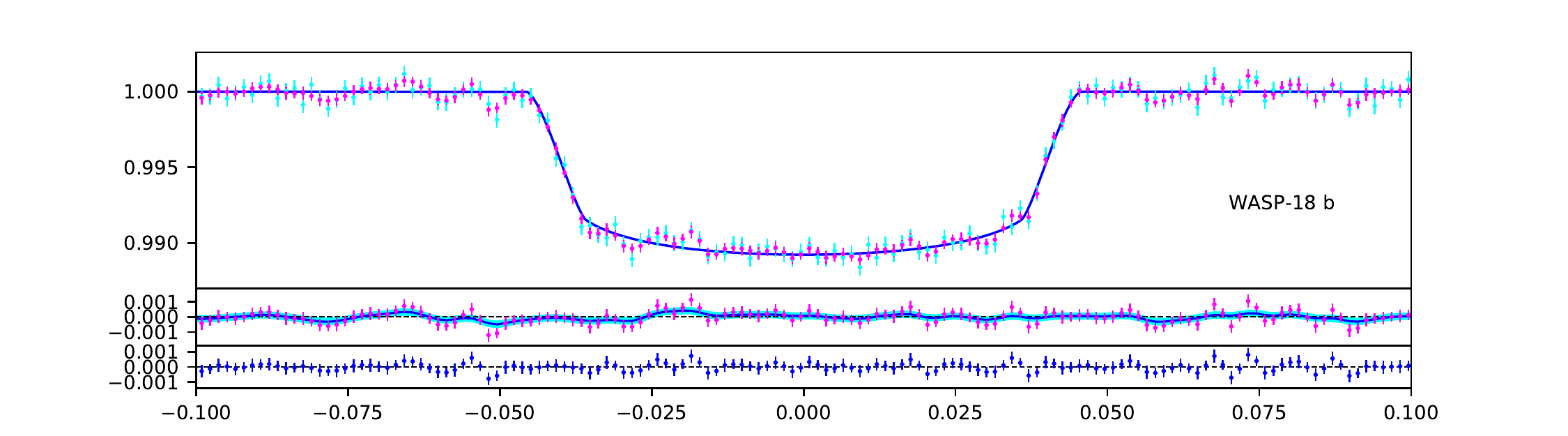}
	\includegraphics[width=0.9\linewidth]{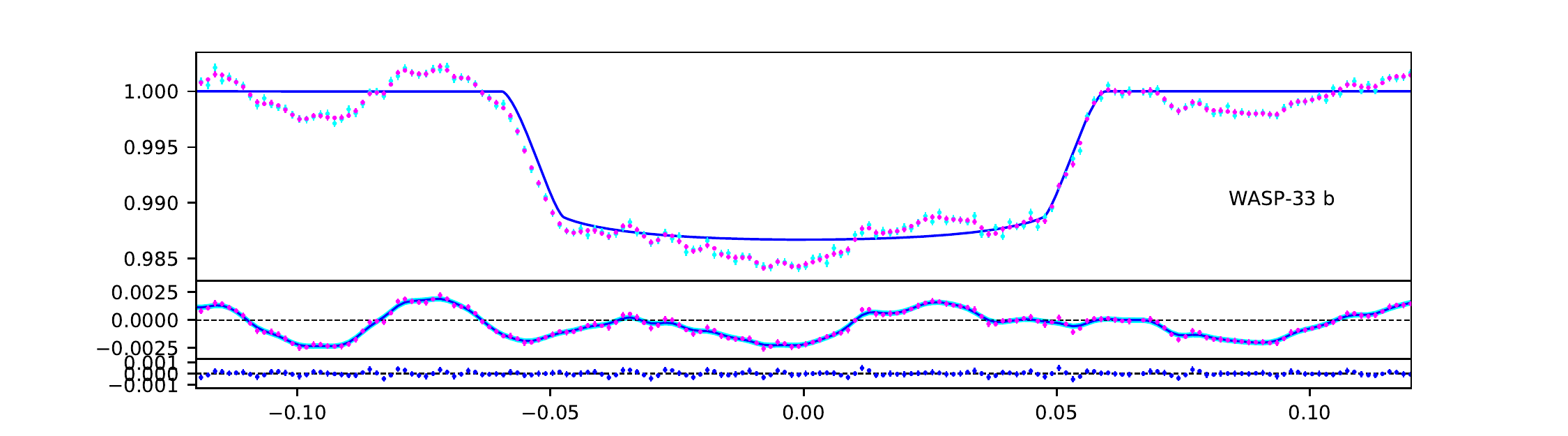}
	\includegraphics[width=0.9\linewidth]{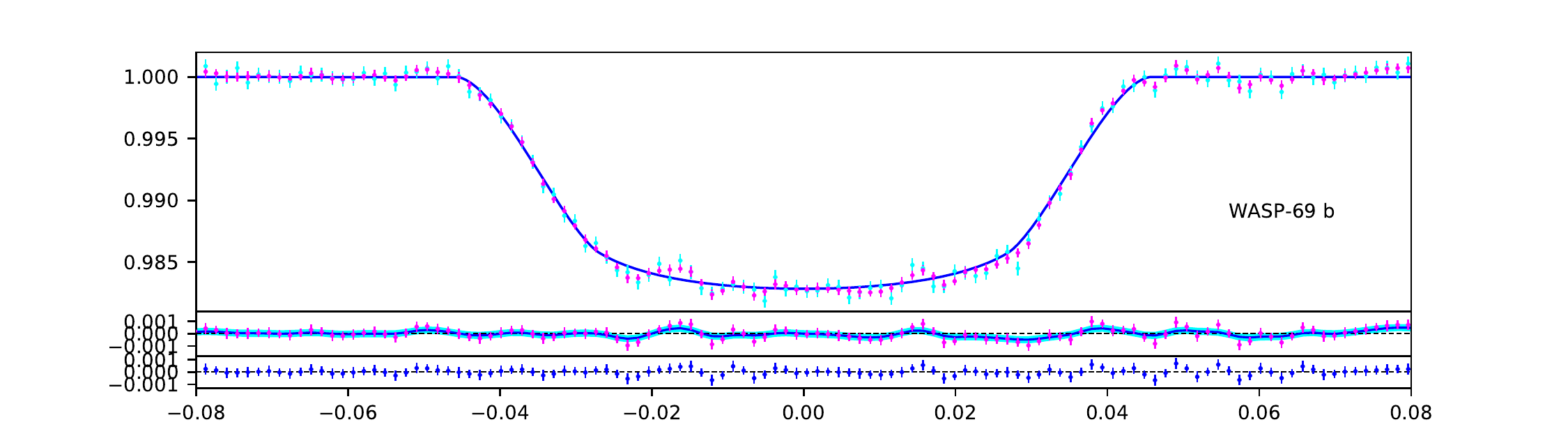}
	\includegraphics[width=0.9\linewidth]{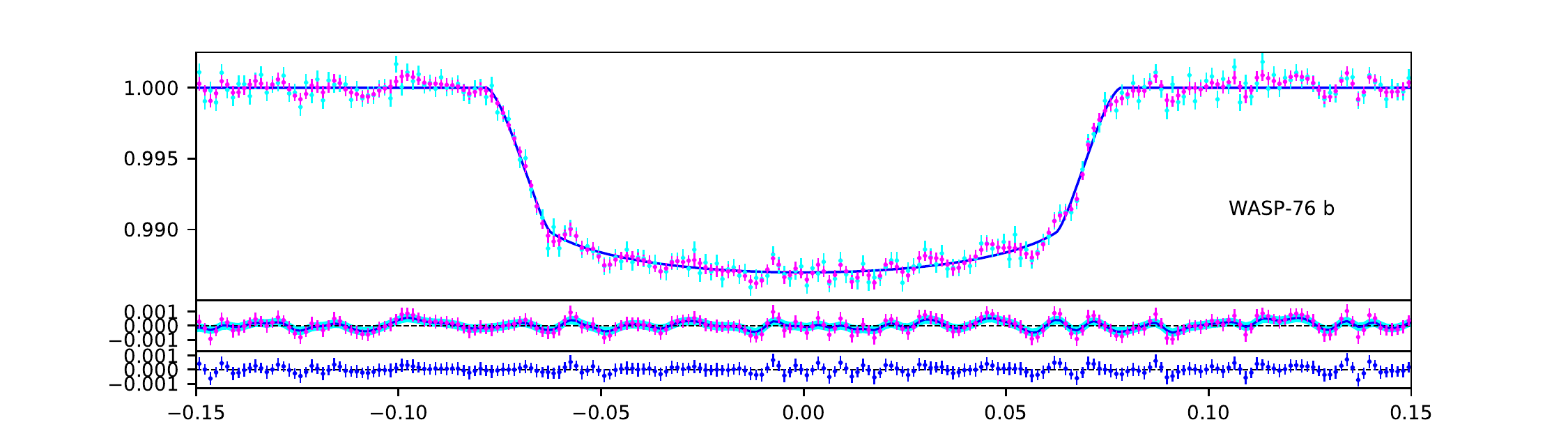}
	\includegraphics[width=0.9\linewidth]{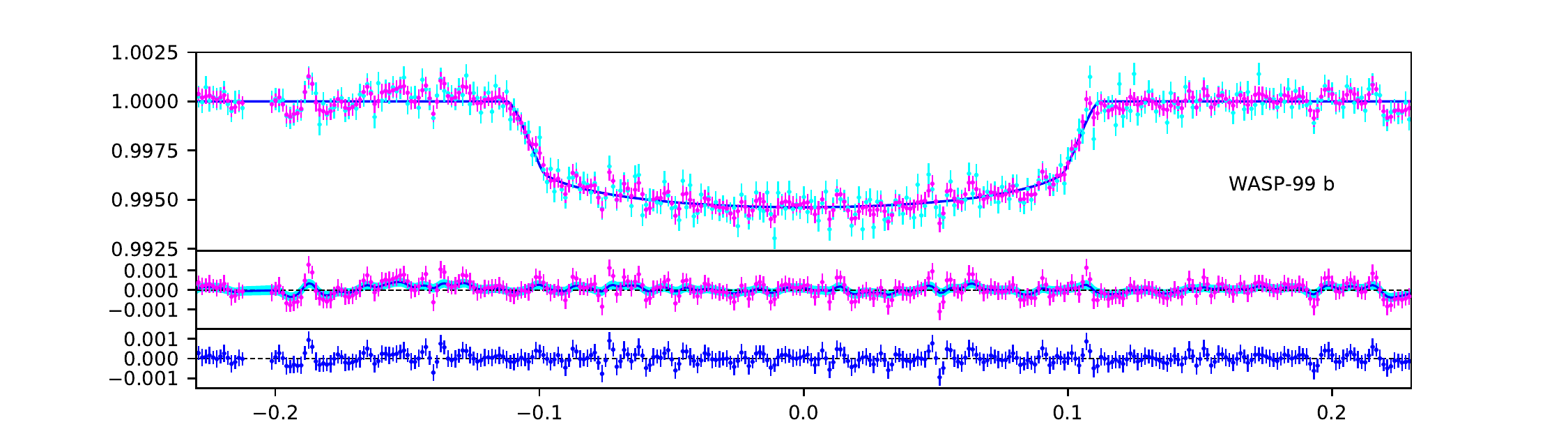}
	\includegraphics[width=0.9\linewidth]{f0.pdf}
	\caption{Same as Figure \ref{fig:fig1}, but for WASP-18 b, WASP-33 b, WASP-69 b, WASP-76 b and WASP-99 b.}
	\label{fig:fig5}
\end{figure*}

\begin{figure*}
	\centering
	\includegraphics[width=0.9\linewidth]{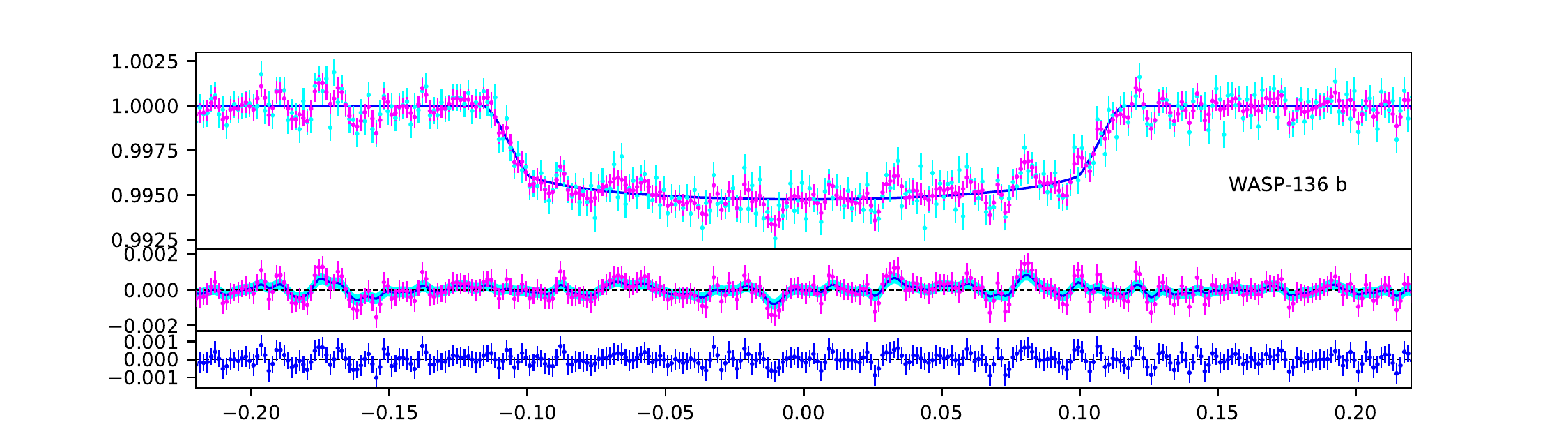}
	\includegraphics[width=0.9\linewidth]{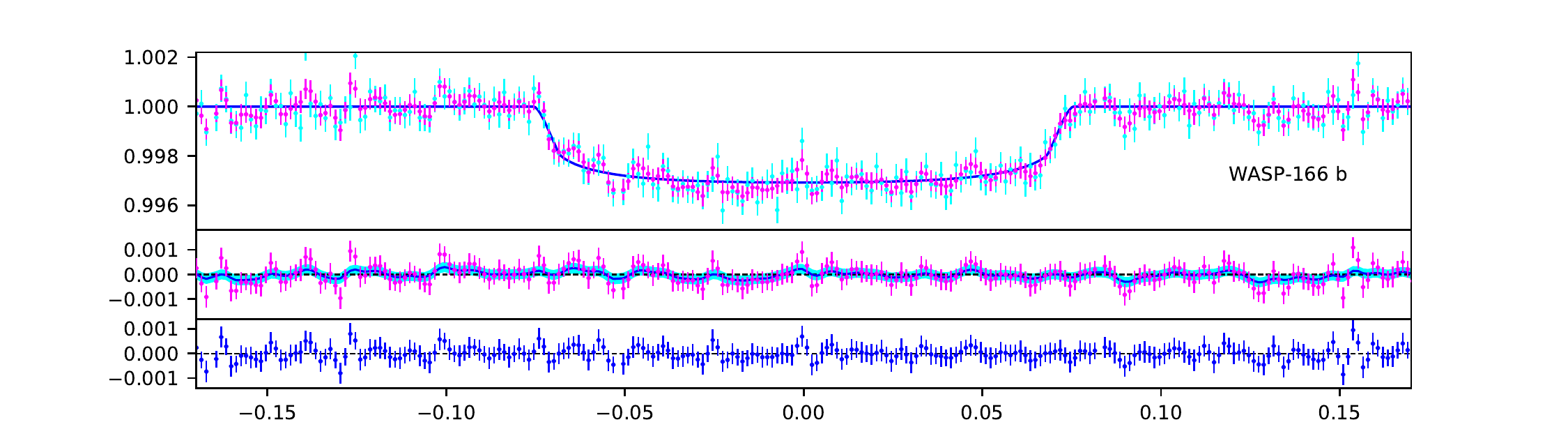}
	\includegraphics[width=0.9\linewidth]{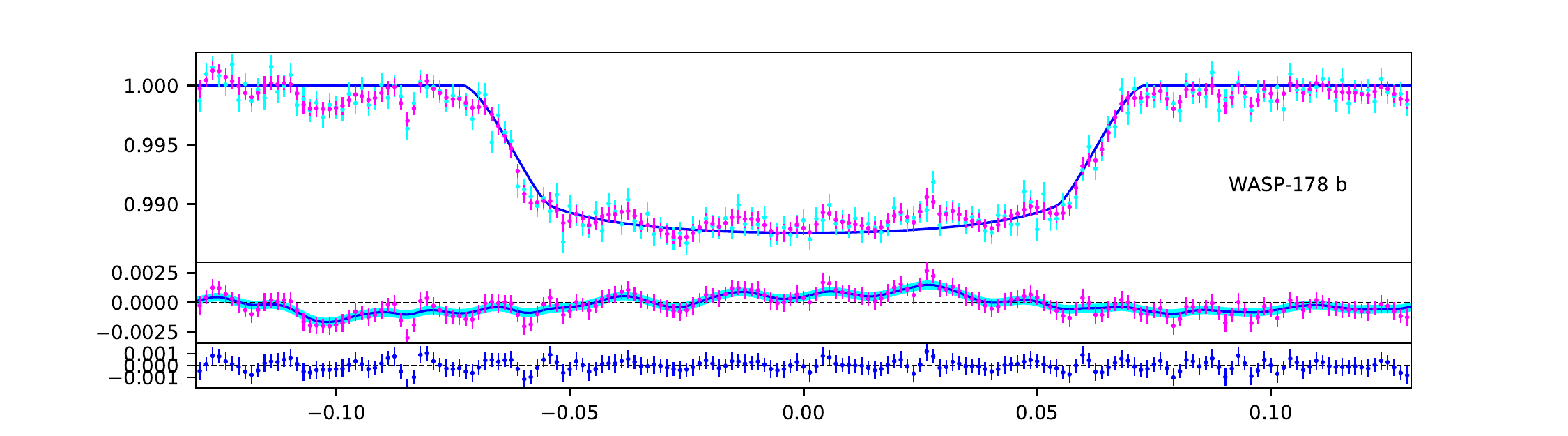}
	\includegraphics[width=0.9\linewidth]{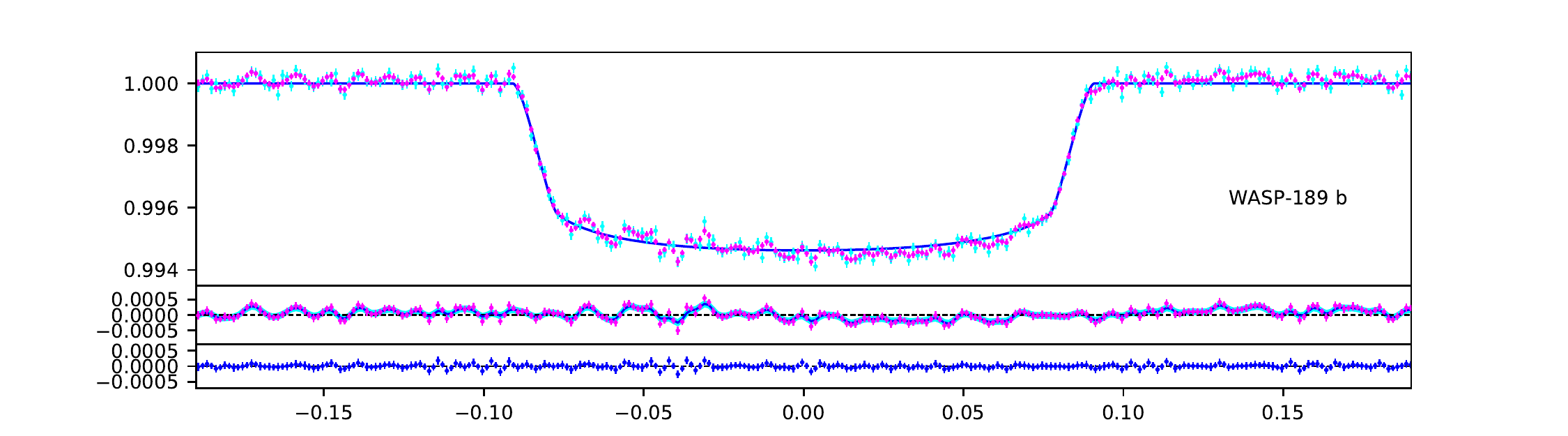}
	\includegraphics[width=0.9\linewidth]{f0.pdf}
	\caption{Same as Figure \ref{fig:fig1}, but for WASP-136 b, WASP-166 b, WASP-178 b and WASP-189 b.}
	\label{fig:fig6}
\end{figure*}

\begin{deluxetable*}{LCCCC}
	\tablecaption{Estimated physical parameters for KELT-2 A b, KELT-3 b, KELT-4 A b and KELT-11 b \label{tab:tab2}}
	\tablewidth{0pt}
	\tablehead{\colhead{Parameter} & \colhead{KELT-2 A b} & \colhead{KELT-3 b} & \colhead{KELT-4 A b} & \colhead{KELT-11 b}}
	\startdata
	\text{Transit parameters}\\
	T_0\; [BJD_{TDB}] & 2459475.42641_{-0.00016}^{+0.00017} & 2458872.85409_{-0.00013}^{+0.00012} & 2459610.39041\pm0.00022 & 2458549.07762_{-0.00067}^{+0.00069}\\
	P\; [days] & 4.113786_{-0.000015}^{+0.000016} & 2.70339033\pm0.00000065 & 2.989582\pm0.000044 & 4.73617\pm0.00027\\
	b & 0.299_{-0.053}^{+0.043} & 0.671_{-0.011}^{+0.01} & 0.63_{-0.016}^{+0.017} & 0.437_{-0.093}^{+0.071}\\
	R_\star/a & 0.1562_{-0.0023}^{+0.0022} & 0.1756\pm0.002 & 0.1602_{-0.0022}^{+0.0024} & 0.2047_{-0.0082}^{+0.0083}\\
	R_p/R_\star & 0.06833\pm0.00023 & 0.09495_{-0.00042}^{+0.00038} & 0.10698_{-0.00082}^{+0.00062} & 0.04644\pm0.00051\\
	\text{Limb-darkening coefficients}\\
	C_1 & 0.323_{-0.051}^{+0.028} & 0.308_{-0.099}^{+0.058} & 0.275_{-0.164}^{++0.094} & 0.471_{-0.057}^{+0.053}\\
	C_2 & 0.079_{-0.047}^{+0.091} & 0.111_{-0.078}^{+0.13} & 0.2_{-0.13}^{+0.25} & 0.091_{-0.068}^{+0.085}\\
	\text{Derived parameters}\\
	T_{14}\; [hr] & 5.061\pm0.012 & 3.171\pm0.011 & 3.259\pm0.016 & 7.112_{-0.038}^{+0.04}\\
	a/R_\star & 6.403_{-0.09}^{+0.097} & 5.696_{-0.064}^{+0.066} & 6.242_{-0.094}^{+0.089} & 4.89_{-0.19}^{+0.2}\\
	i\; [deg] & 87.33_{-0.43}^{+0.51} & 83.23_{-0.18}^{+0.19} & 84.2_{-0.25}^{+0.22} & 84.9_{-1.1}^{+1.3}\\
	M_p\; [M_J] & 1.52_{-0.075}^{+0.077} & 1.479\pm0.063 & 0.899_{-0.058}^{+0.059} & 0.195\pm0.018\\
	M_p\; [M_\earth] & 483\pm24 & 470\pm20 & 286_{-18}^{+19} & 62.1\pm5.9\\
	T_{eq}\; [K] & 1719_{-18}^{+19} & 1868\pm18 & 1756\pm25 & 1717_{-37}^{+39}\\
	a\; [AU] & 0.0542\pm0.0022 & 0.039\pm0.0018 & 0.0465\pm0.0013 & 0.06175_{-0.0045}^{+0.0046}\\
	R_p\; [R_J] & 1.12_{-0.047}^{+0.046} & 1.36\pm0.06 & 1.668\pm0.041 & 1.229\pm0.078\\
	R_p\; [R_\earth] & 13.57\pm0.52 & 15.24\pm0.67 & 18.69\pm0.46 & 13.78\pm0.87\\
	\enddata
\end{deluxetable*}

\begin{deluxetable*}{LCCCC}
	\tablecaption{Estimated physical parameters for KELT-17 b, KELT-19 A b, KELT-20 b and KELT-24 b \label{tab:tab3}}
	\tablewidth{0pt}
	\tablehead{\colhead{Parameter} & \colhead{KELT-17 b} & \colhead{KELT-19 A b} & \colhead{KELT-20 b} & \colhead{KELT-24 b}}
	\startdata
	\text{Transit parameters}\\
	T_0\; [BJD_{TDB}] & 2459502.3949\pm0.00013 & 2458494.13537_{-0.00033}^{+0.00034} & 2458684.314347_{-0.000078}^{+0.000077} & 2458684.816347\pm0.000063\\
	P\; [days] & 3.0801805_{-0.0000093}^{+0.0000094} & 4.61188\pm0.00013 & 3.47410034\pm0.00000034 & 5.55149347\pm0.00000055\\
	b & 0.586\pm0.01 & 0.382_{-0.085}^{+0.066} & 0.5193_{-0.0058}^{+0.006} & 0.096_{-0.059}^{+0.053}\\
	R_\star/a & 0.1587_{-0.0012}^{+0.0013} & 0.1156_{-0.0034}^{+0.0035} & 0.13428_{-0.00048}^{+0.00051} & 0.09355_{-0.00034}^{+0.00058}\\
	R_p/R_\star & 0.09174_{-0.00031}^{+0.00028} & 0.09645_{-0.00083}^{+0.00092} & 0.1157_{-0.00017}^{+0.00016} & 0.087_{-0.00011}^{+0.00013}\\
	\text{Limb-darkening coefficients}\\
	C_1 & 0.287_{-0.079}^{+0.047} & 0.196_{-0.109}^{+0.089} & 0.271_{-0.029}^{+0.031} & 0.263_{-0.015}^{+0.018}\\
	C_2 & 0.1_{-0.066}^{+0.114} & 0.28_{-0.16}^{+0.21} & 0.094_{-0.048}^{+0.046} & 0.165_{-0.033}^{+0.029}\\
	\text{Derived parameters}\\
	T_{14}\; [hr] & 3.4676_{-0.0076}^{+0.0077} & 4.197\pm0.027 & 3.538\pm0.004 & 4.3038_{-0.0042}^{+0.0045}\\
	a/R_\star & 6.302_{-0.051}^{+0.049} & 8.65\pm0.26 & 7.447_{-0.028}^{+0.027} & 10.689_{-0.066}^{+0.039}\\
	i\; [deg] & 84.66\pm0.13 & 87.47_{-0.53}^{+0.62} & 86.001_{-0.061}^{+0.058} & 89.49_{-0.29}^{+0.31}\\
	M_p\; [M_J] & 1.3\pm0.28 & 3.98_{-0.33}^{+0.32} & 3.355_{-0.063}^{+0.062} & 4.65\pm0.16\\
	M_p\; [M_\earth] & 415_{-89}^{+90} & 1265_{-106}^{+101} & 1129\pm20 & 1478_{-50}^{+51}\\
	T_{eq}\; [K] & 2100\pm16 & 1802_{-36}^{+37} & 2329\pm24 & 1408\pm11\\
	a\; [AU] & 0.0482_{-0.0016}^{+0.0017} & 0.0736\pm0.0044 & 0.056\pm0.0014 & 0.0748_{-0.0012}^{+0.0011}\\
	R_p\; [R_J] & 1.469\pm0.049 & 1.717_{-0.093}^{+0.094} & 1.821\pm0.045 & 1.275\pm0.019\\
	R_p\; [R_\earth] & 16.46\pm0.55 & 19.2\pm1 & 20.41_{-0.5}^{+0.51} & 14.29\pm0.21\\
	\enddata
\end{deluxetable*}

\begin{deluxetable*}{LCCCC}
	\tablecaption{Estimated physical parameters for HAT-P-1 b, HAT-P-2 b, HAT-P-11 b and HAT-P-22 b \label{tab:tab4}}
	\tablewidth{0pt}
	\tablehead{\colhead{Parameter} & \colhead{HAT-P-1 b} & \colhead{HAT-P-2 b} & \colhead{HAT-P-11 b} & \colhead{HAT-P-22 b}}
	\startdata
	\text{Transit parameters}\\
	T_0\; [BJD_{TDB}] & 2459829.47609_{-0.00049}^{+0.0005} & 2458956.23792\pm0.00013 & 2458687.20645_{-0.00014}^{+0.00015} & 2458871.629976\pm0.000089\\
	P\; [days] & 4.46508_{-0.00015}^{+0.00016} & 5.6334665\pm0.0000014 & 4.88780248\pm0.00000081 & 3.21223293\pm0.00000058\\
	b & 0.733_{-0.029}^{+0.021} & 0.456_{-0.034}^{+0.027} & 0.107_{-0.081}^{+0.071} & 0.441_{-0.028}^{+0.022}\\
	R_\star/a & 0.1_{-0.0027}^{+0.0028} & 0.1025_{-0.0017}^{+0.0015} & 0.05968_{-0.00035}^{+0.00055} & 0.114_{-0.0014}^{+0.0012}\\
	R_p/R_\star & 0.1161_{-0.0015}^{+0.0013} & 0.06967_{-0.00031}^{+0.00026} & 0.05885_{-0.0003}^{+0.00024} & 0.11019_{-0.00066}^{+0.0005}\\
	\text{Limb-darkening coefficients}\\
	C_1 & 0.22\pm0.15 & 0.301_{-0.05}^{+0.046} & 0.488_{-0.06}^{+0.037} & 0.403_{-0.065}^{+0.056}\\
	C_2 & 0.36_{-0.21}^{+0.22} & 0.099_{-0.075}^{+0.077} & 0.094_{-0.059}^{+0.112} & 0.147_{-0.099}^{+0.116}\\
	\text{Derived parameters}\\
	T_{14}\; [hr] & 2.886_{-0.037}^{+0.04} & 4.2778_{-0.0099}^{+0.0102} & 2.3479_{-0.0061}^{+0.0067} & 2.8598_{-0.009}^{+0.0087}\\
	a/R_\star & 10_{-0.27}^{+0.28} & 9.76_{-0.14}^{+0.17} & 16.756_{-0.154}^{+0.098} & 8.774_{-0.094}^{+0.113}\\
	i\; [deg] & 85.8_{-0.24}^{+0.28} & 87.32_{-0.2}^{+0.24} & 89.63_{-0.25}^{+0.28} & 87.12_{-0.18}^{+0.22}\\
	M_p\; [M_J] & 0.525\pm0.019 & 10.1_{-0.16}^{+0.15} & 0.224\pm0.01 & 2.17_{-0.056}^{+0.055}\\
	M_p\; [M_\earth] & 166.9_{-5.9}^{+6} & 3211_{-50}^{+49} & 71.2\pm3.2 & 690_{-18}^{+17}\\
	T_{eq}\; [K] & 1337_{-22}^{+21} & 1452\pm16 & 826.3_{-9.1}^{+9.2} & 1268\pm14\\
	a\; [AU] & 0.0546_{-0.0019}^{+0.002} & 0.0777_{-0.0029}^{+0.003} & 0.05316_{-0.00081}^{+0.00079} & 0.0425_{-0.0018}^{+0.0019}\\
	R_p\; [R_J] & 1.326_{-0.035}^{+0.034} & 1.159\pm0.041 & 0.391\pm0.0054 & 1.115\pm0.047\\
	R_p\; [R_\earth] & 14.86_{-0.39}^{+0.38} & 12.99\pm0.46 & 4.383\pm0.061 & 12.5_{-0.53}^{+0.52}\\
	\enddata
\end{deluxetable*}

\begin{deluxetable*}{LCCCC}
	\tablecaption{Estimated physical parameters for HAT-P-69 b, HAT-P-70 b, MASCARA-4 b and XO-3 b \label{tab:tab5}}
	\tablewidth{0pt}
	\tablehead{\colhead{Parameter} & \colhead{HAT-P-69 b} & \colhead{HAT-P-70 b} & \colhead{MASCARA-4 b} & \colhead{XO-3 b}}
	\startdata
	\text{Transit parameters}\\
	T_0\; [BJD_{TDB}] & 2459232.985\pm0.0005 & 2459175.05307_{-0.00036}^{+0.00037} & 2459282.43841\pm0.0001 & 2458819.06409\pm0.00026\\
	P\; [days] & 4.78689\pm0.00018 & 2.744219\pm0.000065 & 2.8240776_{-0.0000059}^{+0.0000061} & 3.191585_{-0.000072}^{+0.000071}\\
	b & 0.23_{-0.16}^{+0.12} & 0.554_{-0.043}^{+0.03} & 0.396_{-0.019}^{+0.018} & 0.694_{-0.029}^{+0.027}\\
	R_\star/a & 0.1292_{-0.0028}^{+0.0047} & 0.1837_{-0.0051}^{+0.004} & 0.1802_{-0.15}^{+0.14} & 0.143\pm0.0044\\
	R_p/R_\star & 0.08453_{-0.0007}^{+0.00073} & 0.0924_{-0.00088}^{+0.00075} & 0.08737_{-0.00022}^{+0.0002} & 0.08826_{-0.00069}^{+0.00075}\\
	\text{Limb-darkening coefficients}\\
	C_1 & 0.251_{-0.093}^{+0.08} & 0.409_{-0.105}^{+0.07} & 0.428_{-0.024}^{+0.017} & 0.14_{-0.093}^{+0.104}\\
	C_2 & 0.19_{-0.13}^{+0.18} & 0.14_{-0.1}^{+0.15} & 0.035_{-0.024}^{+0.036} & 0.37\pm0.14\\
	\text{Derived parameters}\\
	T_{14}\; [hr] & 5.027_{-0.028}^{+0.031} & 3.661_{-0.12}^{+0.15} & 3.97_{-0.0074}^{+0.0072} & 2.944_{-0.022}^{+0.021}\\
	a/R_\star & 7.74_{-0.27}^{+0.17} & 5.44_{-0.12}^{+0.15} & 5.549_{-0.044}^{+0.045} & 6.99_{-0.21}^{+0.22}\\
	i\; [deg] & 88.29_{-0.99}^{+1.19} & 84.15_{-0.44}^{+0.6} & 85.91\pm0.22 & 84.31\pm0.4\\
	M_p\; [M_J] & 3.58\pm0.57 & 6.87\pm0.025 & 3.15_{-0.91}^{+0.9} & 12.16\pm0.44\\
	M_p\; [M_\earth] & 1137_{-182}^{+180} & 2183.4_{-7.9}^{+7.8} & 1000_{-288}^{+287} & 3865_{-141}^{+139}\\
	T_{eq}\; [K] & 1953_{-41}^{+45} & 2643_{-45}^{+43} & 2455_{-40}^{+41} & 1719_{-30}^{+29}\\
	a\; [AU] & 0.0692_{-0.0025}^{+0.002} & 0.0499\pm0.002 & 0.0474\pm0.0015 & 0.0448_{-0.0029}^{+0.003}\\
	R_p\; [R_J] & 1.584\pm0.029 & 1.771_{-0.059}^{+0.06} & 1.561\pm0.048 & 1.183_{-0.071}^{+0.072}\\
	R_p\; [R_\earth] & 17.76\pm0.32 & 19.85\pm0.67 & 17.5_{-0.53}^{+0.54} & 13.26\pm0.8\\
	\enddata
\end{deluxetable*}

\begin{deluxetable*}{LCCCC}
	\tablecaption{Estimated physical parameters for WASP-7 b, WASP-8 b, WASP-14 b and WASP-18 b \label{tab:tab6}}
	\tablewidth{0pt}
	\tablehead{\colhead{Parameter} & \colhead{WASP-7 b} & \colhead{WASP-8 b} & \colhead{WASP-14 b} & \colhead{WASP-18 b}}
	\startdata
	\text{Transit parameters}\\
	T_0\; [BJD_{TDB}] & 2459038.75846_{-0.00038}^{+0.00037} & 2458358.92063_{-0.00027}^{+0.00026} & 2459671.35858\pm0.00022 & 2458354.457864_{-0.000048}^{+0.000047}\\
	P\; [days] & 4.95468_{-0.00014}^{+0.00015} & 8.1587277\pm0.000004 & 2.243832_{-0.000046}^{+0.000045} & 0.941452531\pm0.000000085\\
	b & 0.532_{-0.049}^{+0.046} & 0.64\pm0.018 & 0.545_{-0.052}^{+0.033} & 0.395_{-0.022}^{+0.019}\\
	R_\star/a & 0.109_{-0.0033}^{+0.0037} & 0.0777_{-0.0012}^{+0.0013} & 0.1731_{-0.0051}^{+0.004} & 0.2907_{-0.0023}^{+0.0021}\\
	R_p/R_\star & 0.07892_{-0.00065}^{+0.00063} & 0.11747_{-0.00102}^{+0.00087} & 0.09435_{-0.00121}^{+0.00083} & 0.09836_{-0.00032}^{+0.0003}\\
	\text{Limb-darkening coefficients}\\
	C_1 & 0.17\pm0.11 & 0.3_{-0.17}^{+0.12} & 0.122_{-0.088}^{+0.103} & 0.282_{-0.048}^{+0.045}\\
	C_2 & 0.29_{-0.16}^{+0.18} & 0.26_{-0.18}^{+0.25} & 0.5_{-0.18}^{+0.16} & 0.189_{-0.081}^{+0.081}\\
	\text{Derived parameters}\\
	T_{14}\; [hr] & 3.885\pm0.027 & 4.449_{-0.023}^{+0.025} & 2.842\pm0.017 & 2.1907\pm0.0047\\
	a/R_\star & 9.17_{-0.3}^{+0.29} & 12.87_{-0.21}^{+0.19} & 5.78_{-0.13}^{+0.18} & 3.44_{-0.025}^{+0.028}\\
	i\; [deg] & 86.67_{-0.41}^{+0.39} & 87.15_{-0.13}^{+0.12} & 84.58_{-0.45}^{+0.67} & 83.4_{-0.36}^{+0.41}\\
	M_p\; [M_J] & 1.123_{-0.083}^{+0.084} & 2.247_{-0.08}^{+0.078} & 7.3_{-0.49}^{+0.48} & 10.48_{-0.33}^{+0.32}\\
	M_p\; [M_\earth] & 356_{-26}^{+27} & 714\pm25 & 2319_{-156}^{+151} & 3332_{-106}^{+103}\\
	T_{eq}\; [K] & 1522_{-29}^{+30} & 1104\pm18 & 1903_{-40}^{+39} & 2452_{-20}^{+21}\\
	a\; [AU] & 0.0626_{-0.0031}^{+0.0032} & 0.0616\pm0.0026 & 0.0357_{-0.0018}^{+0.0019} & 0.021105\pm0.00098\\
	R_p\; [R_J] & 1.128\pm0.045 & 1.177_{-0.046}^{+0.047} & 1.216\pm0.054 & 1.263\pm0.058\\
	R_p\; [R_\earth] & 12.64\pm0.5 & 13.19_{-0.51}^{+0.52} & 13.63_{-0.61}^{+0.6} & 14.15\pm0.65\\
	\enddata
\end{deluxetable*}

\begin{deluxetable*}{LCCCC}
	\tablecaption{Estimated physical parameters for WASP-33 b, WASP-69 b, WASP-76 b and WASP-99 b \label{tab:tab7}}
	\tablewidth{0pt}
	\tablehead{\colhead{Parameter} & \colhead{WASP-33 b} & \colhead{WASP-69 b} & \colhead{WASP-76 b} & \colhead{WASP-99 b}}
	\startdata
	\text{Transit parameters}\\
	T_0\; [BJD_{TDB}] & 2458792.63408\pm0.00014 & 2459798.77552\pm0.00014 & 2459117.687167_{-0.000072}^{+0.000073} & 2458387.96013\pm0.00027\\
	P\; [days] & 1.219888\pm0.000014 & 3.868143\pm0.000044 & 1.80988122\pm0.00000046 & 5.7525842\pm0.0000025\\
	b & 0.06_{-0.04}^{0.065} & 0.694_{-0.017}^{+0.014} & 0.172_{-0.055}^{+0.04} & 0.109_{-0.069}^{+0.077}\\
	R_\star/a & 0.27236_{-0.00069}^{+0.0013} & 0.08326_{-0.00099}^{+0.00109} & 0.2454\pm0.0017 & 0.11507_{-0.00064}^{+0.00124}\\
	R_p/R_\star & 0.11036_{-0.00064}^{+0.00071} & 0.1271_{-0.0012}^{+0.0011} & 0.10704_{-0.00027}^{+0.00026} & 0.06774_{-0.00022}^{+0.00024}\\
	\text{Limb-darkening coefficients}\\
	C_1 & 0.184_{-0.088}^{+0.073} & 0.124_{-0.089}^{+0.145} & 0.32_{-0.024}^{+0.021} & 0.395_{-0.046}^{+0.04}\\
	C_2 & 0.142_{-0.096}^{+0.124} & 0.59_{-0.23}^{+0.17} & 0.116_{-0.044}^{+0.05} & 0.113_{-0.071}^{+0.084}\\
	\text{Derived parameters}\\
	T_{14}\; [hr] & 2.858_{-0.0056}^{+0.006} & 2.193\pm0.014 & 3.7605_{-0.0051}^{+0.0054} & 5.384\pm0.012\\
	a/R_\star & 3.6716_{-0.0175}^{+0.0093} & 12.01\pm0.15 & 4.075\pm0.028 & 8.691_{-0.093}^{+0.049}\\
	i\; [deg] & 89.07_{-1.02}^{+0.63} & 86.69\pm0.11 & 87.58_{-0.59}^{+0.79} & 89.28_{-0.52}^{+0.45}\\
	M_p\; [M_J] & 1.41_{-0.13}^{+0.14} & 0.259_{-0.017}^{+0.018} & 0.921\pm0.032 & 2.77_{-0.13}^{+0.12}\\
	M_p\; [M_\earth] & 449_{-43}^{+44} & 82.4_{-5.5}^{+5.6} & 293\pm10 & 882\pm40\\
	T_{eq}\; [K] & 1851\pm34 & 959\pm12 & 2190\pm35 & 1484\pm25\\
	a\; [AU] & 0.01364\pm0.00065 & 0.0454_{-0.0016}^{+0.0017} & 0.03277\pm0.00078 & 0.0689\pm0.003\\
	R_p\; [R_J] & 0.858\pm0.041 & 1.006_{-0.036}^{+0.035} & 1.802\pm0.042 & 1.127\pm0.048\\
	R_p\; [R_\earth] & 9.62\pm0.46 & 11.28\pm0.4 & 20.2\pm0.47 & 12.63\pm0.54\\
	\enddata
\end{deluxetable*}

\begin{deluxetable*}{LCCCC}
	\tablecaption{Estimated physical parameters for WASP-136 b, WASP-166 b, WASP-178 b and WASP-189 b \label{tab:tab8}}
	\tablewidth{0pt}
	\tablehead{\colhead{Parameter} & \colhead{WASP-136 b} & \colhead{WASP-166 b} & \colhead{WASP-178 b} & \colhead{WASP-189 b}}
	\startdata
	\text{Transit parameters}\\
	T_0\; [BJD_{TDB}] & 2459092.52455\pm0.00036 & 2458518.96491\pm0.00034 & 2459338.69747_{-0.00026}^{+0.00025} & 2459700.16634_{-0.00018}^{+0.00019}\\
	P\; [days] & 5.215351\pm0.0000079 & 5.4435455\pm0.0000035 & 3.34488\pm0.00013 & 2.724099_{-0.000052}^{+0.000051}\\
	b & 0.331_{-0.103}^{+0.078} & 0.129_{-0.085}^{+0.09} & 0.514_{-0.045}^{+0.035} & 0.358_{-0.052}^{+0.047}\\
	R_\star/a & 0.1369_{-0.004}^{+0.0046} & 0.083382_{-0.001}^{+0.00094} & 0.1388_{-0.0033}^{+0.0029} & 0.2057_{-0.0036}^{+0.004}\\
	R_p/R_\star & 0.0683_{-0.00052}^{+0.00051} & 0.05158_{-0.00041}^{+0.0004} & 0.1068_{-0.0012}^{+0.00092} & 0.06984\pm0.00032\\
	\text{Limb-darkening coefficients}\\
	C_1 & 0.307_{-0.083}^{+0.055} & 0.17_{-0.1}^{+0.14} & 0.24_{-0.11}^{+0.1} & 0.18_{-0.084}^{+0.081}\\
	C_2 & 0.131_{-0.082}^{+0.138} & 0.48_{-0.22}^{+0.22} & 0.24_{-0.17}^{+0.21} & 0.27\pm0.13\\
	\text{Derived parameters}\\
	T_{14}\; [hr] & 5.566_{-0.029}^{+0.032} & 3.61_{-0.02}^{+0.023} & 3.497\pm0.023 & 4.41\pm0.02\\
	a/R_\star & 7.31_{-0.24}^{+0.22} & 11.99_{-0.13}^{+0.15} & 7.2_{-0.15}^{+0.17} & 4.582_{-0.037}^{+0.03}\\
	i\; [deg] & 87.4_{-0.72}^{+0.86} & 89.38_{-0.44}^{+0.41} & 85.9_{-0.36}^{+0.45} & 84.02_{-0.16}^{+0.21}\\
	M_p\; [M_J] & 1.74\pm0.11 & 0.0982\pm0.0051 & 1.66\pm0.12 & 2.01\pm0.15\\
	M_p\; [M_\earth] & 554_{-34}^{+35} & 31.2\pm1.6 & 529\pm39 & 640_{-47}^{+48}\\
	T_{eq}\; [K] & 1638_{-34}^{+36} & 1235\pm12 & 2465\pm47 & 2643\pm28\\
	a\; [AU] & 0.0811\pm0.004 & 0.0699\pm0.0024 & 0.0603_{-0.002}^{+0.0021} & 0.05027\pm0.00074\\
	R_p\; [R_J] & 1.589\pm0.064 & 0.629\pm0.021 & 1.87\pm0.053 & 1.381\pm0.045\\
	R_p\; [R_\earth] & 17.81\pm0.72 & 7.05\pm0.24 & 20.96\pm0.59 & 15.48\pm0.51\\
	\enddata
\end{deluxetable*}

\begin{deluxetable}{lCC}
	\tablecaption{Best-fit Gaussian-process (GP) regression model parameters}
	\label{tab:tab9}
	\tablewidth{0pt}
	\tablehead{\colhead{Target} & \colhead{$\mathrm{\alpha}$} & \colhead{$\mathrm{\tau}$}}
	\startdata
	KELT-2 A b & 0.000191_{-0.000012}^{+0.000013} & 0.00168_{-0.00023}^{+0.00021}\\
	KELT-3 b & 0.000278\pm0.000036 & 0.00379_{-0.00078}^{+0.00152}\\
	KELT-4 A b & 0.000262_{-0.000032}^{+0.000037} & 0.00272_{-0.00034}^{+0.00038}\\
	KELT-11 b & 0.0001753_{-0.000008}^{+0.0000087} & 0.00297_{-0.00021}^{+0.00022}\\
	KELT-17 b & 0.000294_{-0.000023}^{+0.000024} & 0.00289_{-0.00031}^{+0.00029}\\
	KELT-19 A b & 0.000399_{-0.000028}^{+0.000032} & 0.00254_{-0.00024}^{+0.0003}\\
	KELT-20 b & 0.000262_{-0.000011}^{+0.000014} & 0.00275_{-0.00019}^{+0.00023}\\
	KELT-24 b & 0.000191_{-0.000012}^{+0.000013} & 0.00268_{-0.00023}^{+0.00021}\\
	HAT-P-1 b & 0.000849_{-0.000058}^{+0.000062} & 0.00255_{-0.00024}^{+0.0003}\\
	HAT-P-2 b & 0.000221_{-0.000014}^{+0.000013} & 0.00277_{-0.00024}^{+0.00026}\\
	HAT-P-11 b & 0.000325_{-0.000024}^{+0.000023} & 0.00253_{-0.0002}^{+0.00022}\\
	HAT-P-22 b & 0.000248_{-0.000028}^{+0.000031} & 0.00324_{-0.00042}^{+0.00044}\\
	HAT-P-69 b & 0.000394_{-0.000025}^{+0.000023} & 0.003_{-0.00028}^{+0.00029}\\
	HAT-P-70 b & 0.000591_{-0.00003}^{+0.000032} & 0.00267_{-0.0002}^{+0.00023}\\
	MASCARA-4 b & 0.000222_{-0.000014}^{+0.000012} & 0.00289\pm0.00026\\
	XO-3 b & 0.00011_{-0.000063}^{+0.00006} & 0.00436_{-0.0015}^{+0.0031}\\
	WASP-7 b & 0.000305_{-0.000022}^{+0.000019} & 0.00279_{-0.00024}^{+0.00027}\\
	WASP-8 b & 0.00035_{-0.000023}^{+0.000022} & 0.0037_{-0.00027}^{+0.00039}\\
	WASP-14 b & 0.000352_{-0.000027}^{+0.000028} & 0.00268_{-0.00029}^{+0.00037}\\
	WASP-18 b & 0.000251\pm0.000026 & 0.00335_{-0.00063}^{+0.00093}\\
	WASP-33 b & 0.00111_{-0.000095}^{+0.000098} & 0.0127_{-0.0011}^{+0.0014}\\
	WASP-69 b & 0.000302_{-0.00003}^{+0.000032} & 0.00298_{-0.00041}^{+0.00057}\\
	WASP-76 b & 0.00034_{-0.000022}^{+0.000019} & 0.00259_{-0.00022}^{+0.00026}\\
	WASP-99 b & 0.000231\pm0.000018 & 0.00307_{-0.00034}^{+0.00032}\\
	WASP-136 b & 0.000347_{-0.000024}^{+0.000029} & 0.00289_{-0.00028}^{+0.00034}\\
	WASP-166 b & 0.000211_{-0.000019}^{+0.000021} & 0.00295\pm0.00025\\
	WASP-178 b & 0.000662_{-0.000051}^{+0.00006} & 0.00874_{-0.0017}^{+0.0028}\\
	WASP-189 b & 0.0001613_{-0.0000061}^{+0.0000069} & 0.00315_{-0.00017}^{+0.00015}\\
	\enddata
\end{deluxetable}

\begin{figure*}
	\centering
	\includegraphics[width=\linewidth]{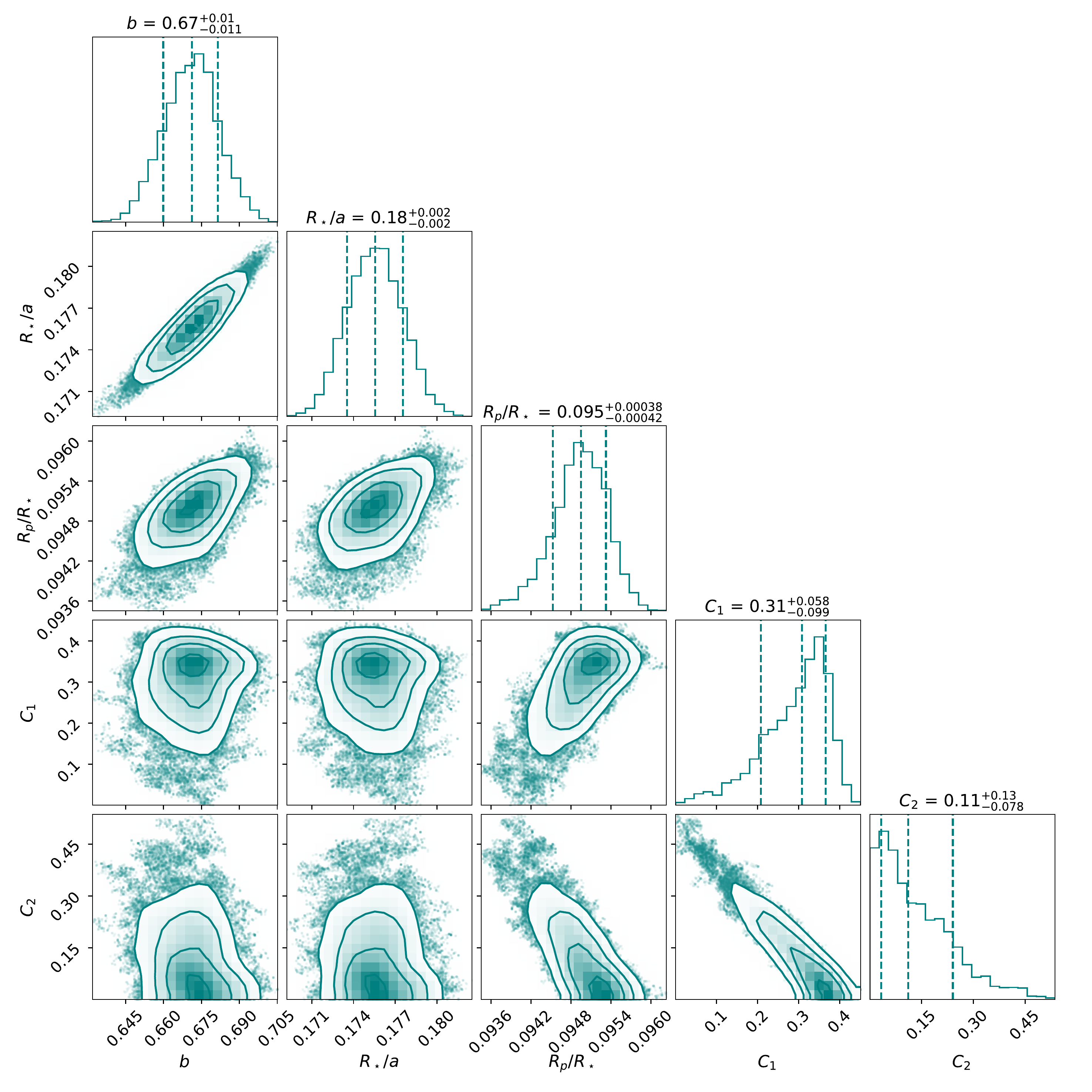}
	\caption{Corner plot showing the posterior distribution of the directly estimated transit parameters from the MCMC sampling for KELT-3 b.}
	\label{fig:fig7}
\end{figure*}

\begin{figure*}
	\centering
	\includegraphics[width=\linewidth]{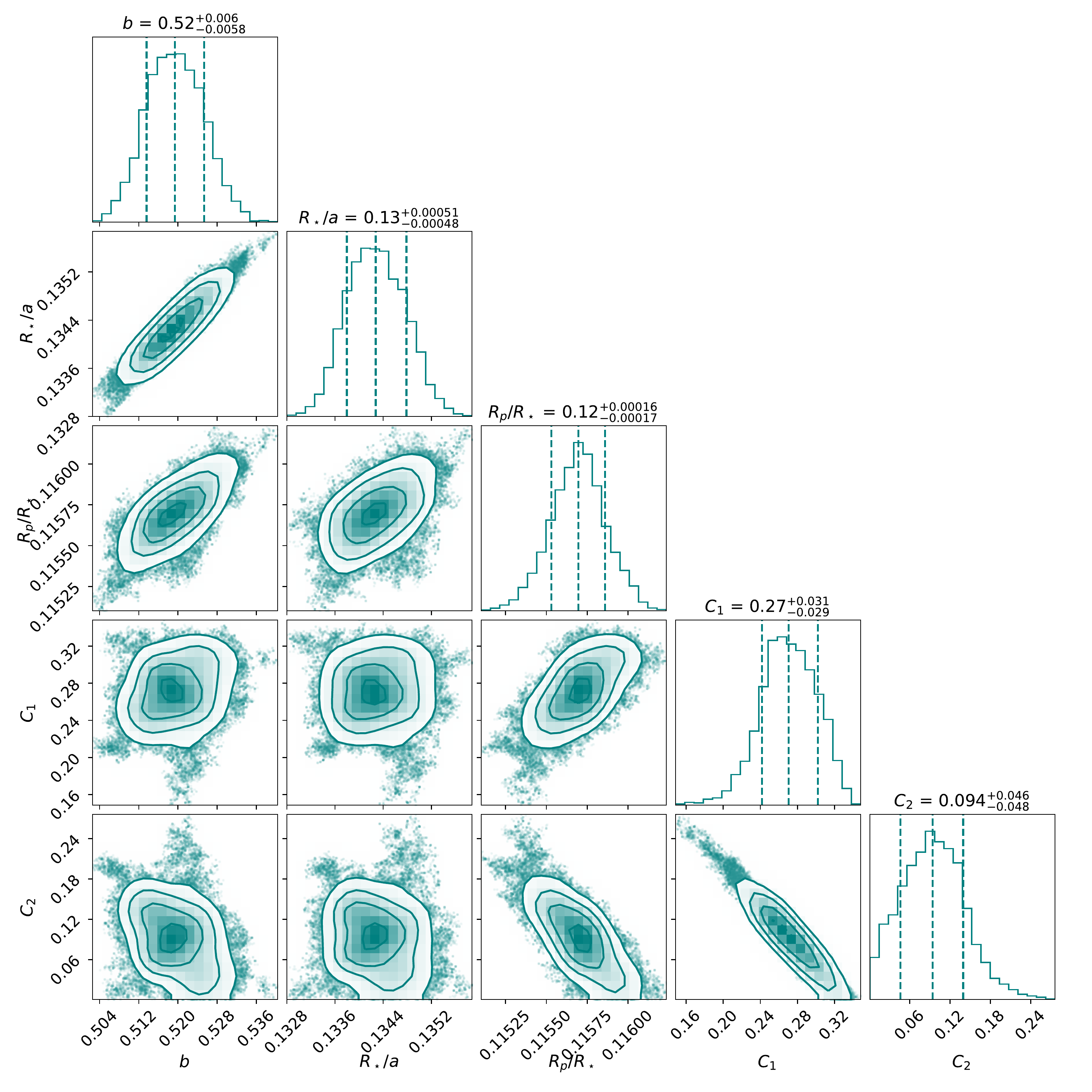}
	\caption{Same as Figure \ref{fig:fig7}, but for KELT-20 b.}
	\label{fig:fig8}
\end{figure*}

\begin{figure*}
	\centering
	\includegraphics[width=\linewidth]{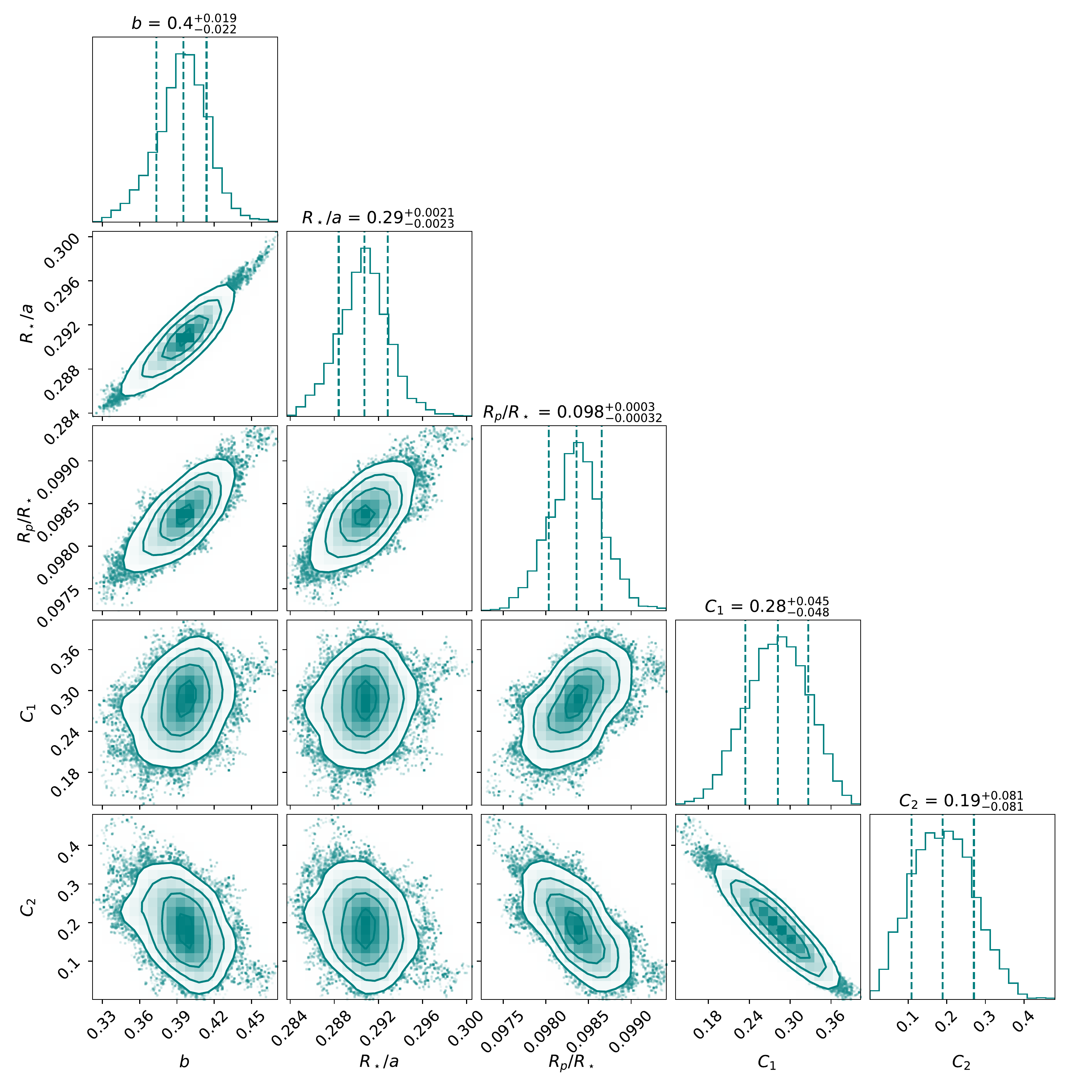}
	\caption{Same as Figure \ref{fig:fig7}, but for WASP-18 b.}
	\label{fig:fig9}
\end{figure*}

\section{Results and Discussions}\label{sec4}

The transit lightcurves corresponding to one transit event for each of the target exoplanets studied in this work are shown in Figures \ref{fig:fig1}-\ref{fig:fig6}. The figures show the unprocessed transit lightcurves from TESS data, the lightcurves after wavelet denoising, the best-fit transit model, the residuals without GP regression along with the best-fit GP regression model, and the final residual flux. It can be noticed from these figures that the wavelet denoising technique has reduced the time uncorrelated fluctuations in the lightcurves without over-smoothing them. This is the advantage of wavelet denoising technique over the traditional techniques, like binning, which also crops out the essential higher frequency signal components from the lightcurves. It can also be seen from the figures that the GP regression technique has modeled the time correlated noise components quite efficiently to render the final residual flux to minimal. One of the major advantage of the GP regression technique is that it can model the correlated noise components with better efficiency for an increase in the SNR of the photometric observations. This is reflected from these figures as well, where the lightcurves with better SNR has the least deviation in the final residual flux.

All the physical properties of the target exoplanets estimated in this study are tabulated in Tables \ref{tab:tab2}-\ref{tab:tab8}. It can be seen from these tables that the precision in the estimated values of the physical properties from this study are reasonably high, owing to the high SNR photometric observations from TESS as well as the implementation of the critical noise treatment algorithm. The corner plots depicting the posterior distribution of the directly estimated transit parameters from the MCMC sampling for KELT-3 b, KELT-20 b and WASP-18 b are shown in Figures \ref{fig:fig7}-\ref{fig:fig9}. These plots show the accuracy in the estimation of the physical properties from this work, and that the uncertainties in the estimated parameters are not underestimated. To understand the extent of improvements in the physical properties for the target exoplanets from this study, the estimated physical properties have been compared with those from the previous studies involving observations from both ground based as well as space based instruments.

Comparing the estimated parameters with the previous studies involving only ground based observations, an improvement in precision in the present study is observed for all the cases. This could be a result of two prime factors. First, the ground based observations are heavily affected by the noise arising from Earth's atmosphere. Hence, even if ground-based telescopes as large as 2-m class have been used in some of the studies, the cumulative noise in those observations might be greater than the smaller space-based TESS observations. The second factor is the lack of implementation of any significant noise reduction technique in the previous studies, which have left the reducible noise components (such as correlated noise) untreated, and have contributed towards the larger uncertainties in the estimated parameters.

On the other hand, comparing the estimated properties with the previous studies involving space based observations would be more interesting. Since all the target exoplanets studied in this work are around bright host stars, several of them have previously been studied using observations from TESS as well as other space based instruments. A comparison between them would give both qualitative and quantitative idea about the capability of different instruments and differences in the results from different approaches. I have compared the three primary transit parameters, $b$, $R_{p}/R_\star$ and $a/R_\star$ from the present work and those from the previous studies, as listed in Table \ref{tab:tab10}. $b$ and $R_{p}/R_\star$ were estimated directly from modeling the transit lightcurves, where as $a/R_\star$ was derived from the directly estimated $R_\star/a$ in the present study.

While comparing with the studies involving larger space based telescopes than TESS, such as Spitzer, HST, Kepler and CHEOPS, a more precise estimation in the transit parameters is expected in the previous studies. However, it has not always been reflected from the comparison detailed as follows. Comparing with \citet{2017AJ....154...25B}, who have studied KELT-11 b using Spitzer observations, their estimated values of $b$ and $a/R_\star$ are more precise than the present study, but that of $R_{p}/R_\star$ is less precise than the present study. \citet{garai2022rapidly} have studied KELT-17 b and KELT-19 A b using CHEOPS and TESS observations. Comparing for KELT-17 b, the estimated values of all three parameters, $b$, $R_{p}/R_\star$ and $a/R_\star$, from their study are less precise than the present study. However, for KELT-19 b, their estimated values of $b$ and $a/R_\star$ are more precise than the present study, but that of $R_{p}/R_\star$ is less precise than the present study. Comparing to \citet{2014MNRAS.437...46N}, who have studied HAT-P-1 b using HST observations, their estimated values of all three parameters, $b$, $R_{p}/R_\star$ and $a/R_\star$, are more precise than the present study. Comparing with \citet{2011ApJ...743...61S}, who have studied HAT-P-11 b using Kepler observations, their estimated value of $b$ and is more precise than the present study, but those of $R_{p}/R_\star$ and $a/R_\star$ are less precise than the present study. Comparing with \citet{2020AA...643A..94L}, who have studied WASP-189 b using CHEOPS observations, their estimated values of all three parameters, $b$, $R_{p}/R_\star$ and $a/R_\star$, are more precise than the present study.

Now, moving towards the previous studies involving observations from TESS, \citet{2022AJ....163..228P} have studied a number of transiting systems, of which KELT-11 b, KELT-20 b, KELT-24 b, HAT-P-2 b, HAT-P-69 b, HAT-P-70 b, XO-3 b, WASP-7 b, WASP-99 b, WASP-136 b and WASP-166 b are common with the present study. By comparing the estimated parameters, except a few cases, such as $b$ and $a/R_\star$ for KELT-11 b, $b$ for HAT-P-69 b, $b$ and $a/R_\star$ for HAT-P-70 b, and $b$ and $a/R_\star$ for WASP-7 b, the precision is better in the present study compared to \citet{2022AJ....163..228P}. \citet{2021AJ....162..263H} have also studied a number of transiting exoplanets, of which KELT-11 b, KELT-19 A b, HAT-P-69 b, HAT-P-70 b, WASP-8 b and WASP-178 b are common with the present study. By comparing the estimated parameters for this case, however, the precision from the present study is better for all the targets compared to the previous work. \citet{2020AcA....70..181M} have studied KELT-24 b using TESS observations, however the precision in the estimated parameters are worse than the present study. \citet{2019AJ....158..141Z} have studied HAT-P-69 b and HAT-p-70 b using TESS observations. Comparing for HAT-P-69 b, the estimated values of $b$ and $a/R_\star$ from \citet{2019AJ....158..141Z} are more precise compared to the present study, while that of $R_{p}/R_\star$ is more precise in the present study. However, comparing for HAT-P-70 b, the estimated values of all three parameters, $b$, $R_{p}/R_\star$ and $a/R_\star$, are more precise in the present study. Comparing to \citet{2014ApJ...794..134W}, who have studied XO-3 b using observations from Spitzer, their estimated values of $R_{p}/R_\star$ and $a/R_\star$ are more precise than the present study. Comparing to \citet{2020AA...636A..98C}, who have studied WASP-18 b, their estimated values for all three parameters, $b$, $R_{p}/R_\star$ and $a/R_\star$, are less precise than the present study. Comparing to \citet{2020AA...639A..34V}, who have studied WASP-33 b using TESS observations, the estimated parameters from the present study are less precise. Comparing to \citet{2020AJ....160..111R}, who have studied WASP-178 b using TESS observations, the estimated values of $b$ and $R_{p}/R_\star$ from the present study are less precise, where as the estimated value of $a/R_\star$ is more precise.

Summarizing the above discussions, the precision of the impact factor, $b$, has improved for KELT-3 b, KELT-4 A b, KELT-17 b, KELT-20 b, KELT-24 b, HAT-P-2 b, HAT-P-22 b, MASCARA-4 b, WASP-8 b, WASP-14 b, WASP-18 b, WASP-69 b, WASP-76 b, WASP-99 b, WASP-136 b and WASP-166 b compared to the previous studies. The improvements have been up to an order of magnitude compared to the most precise values known from the previous studies. Also, the estimated value of $b$ for KELT-2 A b has not been given by any of the previous studies, which is also estimated and updated in the present study.

For the case of $R_p/R_\star$, the precision has improved for KELT-3 b, KELT-4 A b, KELT-11 b, KELT-17 b, KELT-19 A b, KELT-20 b, KELT-24 b, HAT-P-2 b, HAT-P-11 b, HAT-P-22 b, HAT-P-69 b, HAT-P-70 b, MASCARA-4 b, WASP-7 b, WASP-8 b, WASP-18 b, WASP-99 b, WASP-136 b and WASP-166 b compared to the best known values from the previous studies for up to two orders of magnitude. For the cases of KELT-2 A b, WASP-69 b and WASP-76 b, the estimated values of $R_p/R_\star$ has not given by any of the previous studies, which are also estimated and updated in the present study.

For the case of $a/R_\star$, the precision has improved for KELT-2 A b, KELT-3 b, KELT-4 A b, KELT-17 b, KELT-20 b, KELT-24 b, HAT-P-2 b, HAT-P-11 b, HAT-P-22 b, MASCARA-4 b, WASP-8 b, WASP-18 b, WASP-69 b, WASP-99 b, WASP-136 b, WASP-166 b and WASP-178 b compared to the best known values from the previous studies for up to one orders of magnitude. For the case of WASP-76 b, the estimated values of $R_\star/a$ is not given by any of the previous studies, which is also estimated and updated in the present study.

While comparing with the previous studies, it can be noticed that the estimated physical properties from this study vary slightly to significantly compared to the previous studies for most of the cases. While the precision of an estimated parameter depends upon the SNR of the photometric lightcurves and further noise treatments, the accuracy of the values of estimated parameters can still change depending upon the observational and data reduction bias, unidentified noise sources in the lightcurves, insufficient datasets, and inaccurate approaches in data analysis and modeling. For studies involving ground based observations, the chances of observational bias while incorporating various atmospheric factors is quite high. The noise due to various atmospheric perturbations, if unattended, can also contribute to inaccurate estimation of the physical properties. Being a space-based instrument, TESS provides observations unaffected by earth's atmosphere, and hence the data do not contain any correlated or other sources of noise due to the interference of earth's atmosphere. This provides an essential edge to the present study, as the accuracy of the estimated parameters can be considered higher than the previous studies involving only ground based observations.

The next factor that affects the estimated parameters, including the studies involving space based observations, is the unidentified noise components in the observational data. Even the extremely high SNR datasets from space based observations contain correlated noise components originating from short term stellar variability, stellar activities and pulsations. If untreated, these noise components can contribute to inaccurate estimation of the physical parameters, although precision may be high owing to the high SNR data. In the present study, the previous proved and majorly accepted GP regression technique has been used to efficiently model the correlated noise components in the transit lightcurves which modeling for the transit signal. This reduces the affect of these noise components in the estimated parameters from modeling, thus making them more accurate. Since almost all the previous studies of the target exoplanets studied in this work have not adopted any kind of correlated noise treatment technique, the estimated parameters from this work can be treated as more accurate.

\startlongtable
\begin{deluxetable*}{llcCCC}
	\tablecaption{Comparison of estimated parameters with the previous studies involving observations from Spitzer, HST, Kepler, CHEOPS and TESS.}
	\label{tab:tab10}
	\tablewidth{0pt}
	\tablehead{\colhead{Target} & \colhead{Study} & \colhead{Instrument} & \colhead{$b$} & \colhead{$R_{p}/R_\star$} & \colhead{$a/R_\star$}}
	\startdata
	KELT-11 b & This work & & 0.437_{-0.093}^{+0.071} & 0.04644\pm0.00051 & 4.89_{-0.19}^{+0.2}\\
	& \citet{2017AJ....154...25B} & Spitzer & 0.404_{-0.018}^{+0.013} & 0.0514_{-0.0038}^{+0.0032} & 4.98\pm0.05\\
	& \citet{2022AJ....163..228P} & TESS & 0.541_{-0.061}^{+0.044} & 0.0475\pm0.0006 & 4.61_{-0.15}^{+0.18}\\
	& \citet{2021AJ....162..263H} & TESS & 0.488\pm0.074 & 0.04725\pm0.00066 & \nodata\\
	KELT-17 b & This work & & 0.586\pm0.01 & 0.09174_{-0.00031}^{+0.00028} & 6.302_{-0.051}^{+0.049}\\
	& \citet{garai2022rapidly} & CHEOPS, TESS & 0.587\pm0.011 & 0.0921\pm0.0011 & 6.246\pm0.077\\
	KELT-19 A b & This work & & 0.382_{-0.085}^{+0.066} & 0.09645_{-0.00083}^{+0.00092} & 8.65\pm0.26\\
	& \citet{garai2022rapidly} & CHEOPS, TESS & 0.499\pm0.018 & 0.0985\pm0.001 & 8.213\pm0.088\\
	& \citet{2021AJ....162..263H} & TESS & 0.367\pm0.106 & 0.09649\pm0.00115 & \nodata \\
	KELT-20 b & This work & & 0.5193_{-0.0058}^{+0.006} & 0.1157_{-0.00017}^{+0.00016} & 7.447_{-0.028}^{+0.027}\\
	& \citet{2022AJ....163..228P} & TESS & 0.502_{-0.016}^{+0.017} & 0.1157\pm0.0005 & 7.53_{-0.07}^{+0.06}\\
	KELT-24 b & This work & & 0.096_{-0.059}^{+0.053} & 0.087_{-0.00011}^{+0.00013} & 10.689_{-0.066}^{+0.039}\\
	& \citet{2020AcA....70..181M} & TESS & \nodata & 0.0901_{-0.0004}^{+0.0003} & 7.89_{-0.12}^{+0.14}\\
	& \citet{2022AJ....163..228P} & TESS & 0.135_{-0.076}^{+0.06} & 0.0871_{-0.0002}^{+0.0003} & 9.97_{-0.09}^{+0.07}\\
	HAT-P-1 b & This work & & 0.733_{-0.029}^{+0.021} & 0.1161_{-0.0015}^{+0.0013} & 10_{-0.27}^{+0.28}\\
	& \citet{2014MNRAS.437...46N} & HST & 0.7501_{-0.0069}^{+0.0064} & 0.11802\pm0.00018 & 9.853\pm0.071\\
	HAT-P-2 b & This work & & 0.456_{-0.034}^{+0.027} & 0.06967_{-0.00031}^{+0.00026} & 9.76_{-0.14}^{+0.17}\\
	& \citet{2022AJ....163..228P} & TESS & 0.457_{-0.04}^{+0.035} & 0.0691\pm0.0004 & 9.04_{-0.18}^{+0.19}\\
	HAT-P-11 b & This work & & 0.107_{-0.081}^{+0.071} & 0.05885_{-0.0003}^{+0.00024} & 16.756_{-0.154}^{+0.098}\\
	& \citet{2011ApJ...743...61S} & Kepler & 0.132\pm0.045 & 0.05862\pm0.00026 & 15.6\pm1.5\\
	HAT-P-69 b & This work & & 0.23_{-0.16}^{+0.12} & 0.08453_{-0.0007}^{+0.00073} & 7.74_{-0.27}^{+0.17}\\
	&\citet{2019AJ....158..141Z} & TESS & 0.366_{-0.05}^{+0.06} & 0.08703_{-0.0008}^{+0.00075} & 7.32_{-0.18}^{+0.16}\\
	& \citet{2022AJ....163..228P} & TESS & 0.26_{-0.15}^{+0.1} & 0.0849_{-0.0008}^{+0.0009} & 7.68_{-0.25}^{+0.19}\\
	& \citet{2021AJ....162..263H} & TESS & 0.463\pm0.27 & 0.0865\pm0.0121 & \nodata \\
	HAT-P-70 b & This work & & 0.554_{-0.043}^{+0.03} & 0.0924_{-0.00088}^{+0.00075} & 5.44_{-0.12}^{+0.15}\\
	&\citet{2019AJ....158..141Z} & TESS & 0.629_{-0.054}^{+0.081} & 0.09887_{-0.00095}^{+0.00133} & 5.45_{-0.49}^{+0.29}\\
	& \citet{2022AJ....163..228P} & TESS & 0.543_{-0.032}^{+0.028} & 0.093_{-0.0009}^{+0.0008} & 5.52_{-0.11}^{+0.12}\\
	& \citet{2021AJ....162..263H} & TESS & 0.464\pm0.267 & 0.08712\pm0.01055 & \nodata \\
	XO-3 b & This work & & 0.694_{-0.029}^{+0.027} & 0.08826_{-0.00069}^{+0.00075} & 6.99_{-0.21}^{+0.22}\\
	& \citet{2014ApJ...794..134W} & Spitzer & \nodata & 0.08825\pm0.00037 & 7.052_{-0.097}^{+0.076}\\
	& \citet{2022AJ....163..228P} & TESS & 0.696_{-0.033}^{+0.028} & 0.0888\pm0.0011 & 7.09_{-0.23}^{+0.24}\\
	WASP-7 b & This work & & 0.532_{-0.049}^{+0.046} & 0.07892_{-0.00065}^{+0.00063} & 9.17_{-0.3}^{+0.29}\\
	& \citet{2022AJ....163..228P} & TESS & 0.53_{-0.05}^{+0.038} & 0.079_{-0.0006}^{+0.0007} & 8.93_{-0.25}^{+0.29}\\
	WASP-8 b & This work & & 0.64\pm0.018 & 0.11747_{-0.00102}^{+0.00087} & 12.87_{-0.21}^{+0.19}\\
	& \citet{2021AJ....162..263H} & TESS & 0.601\pm0.026 & 0.11925\pm0.00181 & \nodata \\
	WASP-14 b & This work & & 0.545_{-0.052}^{+0.033} & 0.09435_{-0.00121}^{+0.00083} & 5.78_{-0.13}^{+0.18}\\
	& \citet{2015ApJ...811..122W} & Spitzer & \nodata & 0.09419\pm0.00043 & 5.99\pm0.09\\
	WASP-18 b & This work & & 0.395_{-0.022}^{+0.019} & 0.09836_{-0.00032}^{+0.0003} & 3.44_{-0.025}^{+0.028}\\
	& \citet{2020AA...636A..98C} & TESS & 0.36_{-0.18}^{+0.11} & 0.1018\pm0.0011 & 3.48_{-0.17}^{+0.16}\\
	WASP-33 b & This work & & 0.06_{-0.04}^{0.065} & 0.11036_{-0.00064}^{+0.00071} & 3.6716_{-0.0175}^{+0.0093}\\
	& \citet{2020AA...639A..34V} & TESS & \nodata & 0.10716\pm0.00023 & 3.605\pm0.009\\
	WASP-99 b & This work & & 0.109_{-0.069}^{+0.077} & 0.06774_{-0.00022}^{+0.00024} & 8.691_{-0.093}^{+0.049}\\
	& \citet{2022AJ....163..228P} & TESS & 0.1_{-0.065}^{+0.085} & 0.0678\pm0.0003 & 8.71_{-0.1}^{+0.04}\\
	& \citet{2021AJ....162..263H} & TESS & 0.202\pm0.118 & 0.06851\pm0.00058 & \nodata \\
	WASP-136 b & This work & & 0.331_{-0.103}^{+0.078} & 0.0683_{-0.00052}^{+0.00051} & 7.31_{-0.24}^{+0.22}\\
	& \citet{2022AJ....163..228P} & TESS & 0.344_{-0.121}^{+0.080} & 0.0682\pm0.0008 & 7.31_{-0.25}^{+0.26}\\
	WASP-166 b & This work & & 0.129_{-0.085}^{+0.09} & 0.05158_{-0.00041}^{+0.0004} & 11.99_{-0.13}^{+0.15}\\
	& \citet{2022AJ....163..228P} & TESS & 0.365_{-0.089}^{+0.104} & 0.0517\pm0.0009 & 11.25_{-0.52}^{+0.38}\\
	WASP-178 b & This work & & 0.514_{-0.045}^{+0.035} & 0.1068_{-0.0012}^{+0.00092} & 7.2_{-0.15}^{+0.17}\\
	& \citet{2020AJ....160..111R} & TESS & 0.628_{-0.029}^{+0.027} & 0.11066_{-0.00087}^{+0.0009} & 6.49\pm0.18\\
	& \citet{2021AJ....162..263H} & TESS & 0.433\pm0.253 & 0.10538\pm0.0107 & \nodata \\
	WASP-189 b & This work & & 0.358_{-0.052}^{+0.047} & 0.06984\pm0.00032 & 4.582_{-0.037}^{+0.03}\\
	& \citet{2020AA...643A..94L} & CHEOPS & 0.478_{-0.012}^{+0.009} & 0.07045_{-0.00015}^{+0.00013} & 4.6_{-0.025}^{+0.031}\\
	\enddata
\end{deluxetable*}

Another major factor contributing to the inaccurate estimation of physical properties is insufficient datasets. A single lightcurve could contain unidentifiable noise components, which could be treated as a part of the transit signal while modeling the lightcurve. This will contribute to inaccurate estimation of the physical properties. Similar issue can also occur while using multiple but incomplete transit observations. By using multiple full transit observations, such bias in modeling the transit signal can be overcome, resulting in more accurate parameter estimation. Some of the previous studies of the target exoplanets in this work, including some of the past studies involving TESS data, have used very limited datasets, and which could have contributed to some bias in the estimated parameters. On the other hand, the present study has used extensive datasets from TESS observations covering a decent number of full transit observations for each of the target exoplanets, which is expected to have minimized such bias in the estimated properties.

Apart from the above factors, the wavelet dependent parameters, such as $R_p/R_\star$ would vary between studies depending upon the wavelength range of the photometric observations. In such cases, where the previous studies involve observations from a different space based telescope, such as Spitzer, HST or CHEOPS, the estimated values for $R_p/R_\star$ can be considered as complementary to the previously known values, thereby providing the scope for future multi-wavelength studies to characterize the planetary atmospheres.
	
Summarizing up the above discussions, the estimated transit parameters for the target exoplanets resulting from this work can be treated as extremely accurate and precise, and compared to the previous studies of these exoplanets, they can be regarded as the updated parameter values for most of the cases.

The other dependent physical properties, which were derived using the directly estimated transit parameters and the stellar properties adopted from literature, also showed similar trends in improvements in their estimated values, as compared to the previous studies, which is expected. As a result, these parameter values are also extremely accurate and precise, and as such, can be regarded as the updated physical properties for the target exoplanets for most of the cases.

The orbital period, $P$, is not estimated directly from transit modeling of the lightcurves, but from the estimated mid-transit times. $P$ depends upon the total span of time over which the transit observations have been conducted. Since, I have only used the TESS transit photometric observations of the target exoplanets in this study, depending upon the span of the period over which the TESS data is available, the precision in the estimated value of $P$ can vary significantly for each of the cases. For the cases where the target has been observed only in a single sector or a few consecutive sectors, the precision for $P$ is comparatively less irrespective of the number of transits observed over that period. However, for the cases where the target has been observed in at least two sectors separated by a large time scale, the precision in the estimated values of $P$ are very high. Comparing with the previous studies, the precision in the estimated values of $P$ from this study are higher for KELT-20 b, KELT-24 b and WASP-76 b, and almost similar for KELT-2 A b, KELT-3 b, MASCARA-4 b, WASP-99 b and WASP-166 b. For other cases, the precision in the estimated values of $P$ were less compared to at least one of the previous studies, which can be attributed to the shorter total time span of TESS observations as discussed above. However, with more TESS observations of these targets in future sectors, $P$ can be estimated more precisely.

Other than the planetary properties, the quadratic limb darkening coefficients for the host stars were also estimated precisely while modeling the transit lightcurves. They are given along with the planetary properties in Tables \ref{tab:tab2}-\ref{tab:tab8}. The best-fit GP regression model parameters for each of the targets are given in Table \ref{tab:tab9}.

The major output from this study has been the updated physical properties of 28 transiting exoplanets orbiting around bright stars with $V_{mag} \le 10$. These updated parameter values, as discussed above, are more precise for most of the cases compared to the previous studies, as well as more accurate and reliable. In the present era of large scale studies in the field exoplanet science, these updated values of physical properties of several known exoplanets would be immensely useful in a plethora of different studies, starting from the studies of planetary evolution and dynamics \citep[e.g.][]{2020AJ....159..207B, 2022ApJ...931...10R, 2022ApJ...941L..31V, 2022AJ....164...26H, 2020ApJ...902L...5P}, to their compositional studies \citep[e.g.][]{2022AJ....164...15E, 2022ApJ...941..155B, 2022arXiv221101800S, 2020Icar..35214025O} and the search for planetary companions, such as exomoons \citep[e.g.][]{2020MNRAS.499.4195T, 2013MNRAS.432.2994F, 2022ApJ...936....2S, 2022arXiv220611368T} etc. This study also demonstrates how large scale survey missions of future can shape our understanding of existing planetary population even further.

I thank Darin Ragozzine for his valuable suggestions in improving the manuscript. I thank the anonymous reviewer for his/her valuable comments and suggestions. Some of the computational results reported in this work were performed on the high performance computing facility (NOVA) of IIA, Bangalore. I am thankful to the computer division of Indian Institute of Astrophysics for the help and co-operation extended to us. This paper includes data collected by the TESS mission, which are publicly available from the Mikulski Archive for Space Telescopes (MAST). I acknowledge the use of public TOI Release data from pipelines at the TESS Science Office and at the TESS Science Processing Operations Center. Funding for the TESS mission is provided by NASA’s Science Mission directorate. Support for MAST is provided by the NASA Office of Space Science via grant NNX13AC07G and by other grants and contracts. This research made use of Lightkurve, a Python package for Kepler and TESS data analysis.

\bibliography{ms}{}
\bibliographystyle{aasjournal}

\end{document}